\documentstyle[PASJadd,epsf]{PASJ95}

\makeatletter

\long\def\@makecaption#1#2{
 \vskip 10pt 
 \setbox\@tempboxa\hbox{#2}
 \ifdim \wd\@tempboxa >\hsize #2\par \else \hbox
to\hsize{\hfil\box\@tempboxa\hfil} 
 \fi}

\makeatother

\begin{document}

\title{Exploring Galaxy Evolution from Infrared Number Counts and 
Cosmic Infrared Background}

\author{Tsutomu T. {\sc Takeuchi}, \\
  {\it Division of Particle and Astrophysical Sciences, Nagoya University, 
    Chikusa-ku, Nagoya 464--8602}\\
  Takako T. {\sc Ishii}, \\
  {\it Kwasan and Hida observatories, Kyoto University, Yamashina-ku, 
    Kyoto, 607--8471}\\
  Hiroyuki {\sc Hirashita}\thanks{Research Fellows of the Japan Society for
the Promotion of Science.} ,  Kohji {\sc Yoshikawa}$^{*}$,\\
  {\it Department of Astronomy, Kyoto University, Sakyo-ku, Kyoto 
    606--8502}\\
  Hideo {\sc Matsuhara}, \\
  {\it Institute of Space and Astronautical Sciences, Sagamihara, Kanagawa, 
    229--8510}\\
  Kimiaki {\sc Kawara}, \\
  {\it Institute of Astronomy, University of Tokyo, Mitaka, Tokyo, 181--8588}\\
  and\\
  Haruyuki {\sc Okuda}\\
  {\it Gunma Astronomical Observatory, Nakayama, Takayama, Agatsuma, 
    Gunma, 377--0702}\\
  {\it E-mail(TTT): takeuchi@u.phys.nagoya-u.ac.jp}\\
}

\abst{
Recently reported infrared (IR) galaxy number counts and cosmic infrared 
background (CIRB) all suggest that galaxies have experienced a strong 
evolution sometime in their lifetime.
We statistically estimate the galaxy evolution history from these data.
The evolution of galaxy luminosity is treated as a nonparametric stepwise 
function, in order to explore the most suitable evolution history which 
satisfies the constraint from CIRB.
We find that an order of magnitude increase of the far-infrared (FIR) 
luminosity at redshift $z=0.5 \mbox{--} 1.0$ is necessary 
to reproduce the very high CIRB intensity at $140\; \mu$m reported 
by Hauser et al.\ (1998). 
We note that an evolutionary factor larger than ten at high-$z$ ($z > 2$)
overpredicts the CIRB intensity at submillimeter wavelength regime.
Basically the evolutionary factor increases by a factor of 30 up to 
$z \sim 0.75$ and decreases to, even at most, a factor of 10 toward 
$z \sim 5$, though many variants are allowed within these constraints.
This evolution history also satisfies the constraints from 
the galaxy number counts obtained by {\sl IRAS}, {\sl ISO} and, 
roughly, SCUBA.
The rapid evolution of the comoving IR luminosity density 
required from the CIRB well reproduces the very steep slope of galaxy number 
counts obtained by {\sl ISO}.
This result is also consistent with some recent results of multiwavelength
surveys and follow-up observations of {\sl ISO}~extragalactic sources.
We also estimate the cosmic star formation history (SFH) from the
obtained FIR luminosity density, considering the effect of the metal 
enrichment in galaxies.
The derived SFH increases steeply with redshift in $0 \ltsim z \ltsim 0.75$, 
and becomes flat or even declines at $z > 0.75$.
This is consistent with the SFH estimated from the reported ultraviolet 
luminosity density.
In addition, we present the performance of the Japanese 
ASTRO-F FIR galaxy survey.
We show the expected number counts in the survey.
We also evaluate how large a sky area is necessary to derive a secure 
information of galaxy evolution up to $z \sim 1$ from the survey, 
and find that at least 50 -- 300~${\rm deg^2}$ is required.

}

\kword{Galaxies: active --- Galaxies: evolution --- Galaxies: formation --- 
Galaxies: starburst --- Infrared: galaxies}

\maketitle
\thispagestyle{headings}

\section{INTRODUCTION}

The galaxy evolution has long been a strong driving force of the
cosmological studies, and many problems have been still unsolved.
Not only optical but also infrared (hereafter IR) and submillimeter 
(sub-mm) waveband observations of galaxies 
have a crucial importance for full understanding of their 
evolutionary status.

Recent infrared and sub-mm surveys revealed very steep
slopes of galaxy number counts compared with that expected
from the no-evolution model, and provided a new 
impetus to the related field (e.g.\ Kawara et~al.~1998; 
Puget et~al.~1999; Dole et~al.~2000; Oliver et~al.~2000a; 
Kawara et~al.~2000; Okuda et~al.~2000).
The 170-$\mu$m slope proved to be ${\rm d}\log N/{\rm d}\log S \ltsim -2.5$ 
at $S\simeq 0.5$ Jy in these new deep surveys, while the slope would be 
$\sim -1.5$ in the no evolution case.
Such an excess of galaxy number count is generally understood 
as a consequence of a strong galaxy evolution, i.e.\ a rapid 
change of the star formation rate in galaxies.

Another important related issue is the cosmic infrared background (CIRB), 
which is the integrated IR light from galaxies, 
especially from those are so faint that are unable to be resolved.
Therefore the CIRB provides an important information on the past star 
formation history of galaxies including the inaccessible sources.
The CIRB has detected by {\sl COBE}~(Puget et~al.~1996; Fixsen et~al.~1998; 
Hauser et~al.~1998).
The reported CIRB intensity was surprisingly high 
($\nu L_\nu = 25 \pm 7$ nW$\,{\rm m^{-2}sr^{-1}}$ at 140 $\mu$m and 
$\nu L_\nu = 14 \pm 3$ nW$\,{\rm m^{-2}sr^{-1}}$ at 240 $\mu$m: 
Hauser et al.~1998), and provided a profound problem awaiting 
to be solved.

\begin{figure}[t]
\epsfxsize=7.0cm
\centerline{\epsfbox{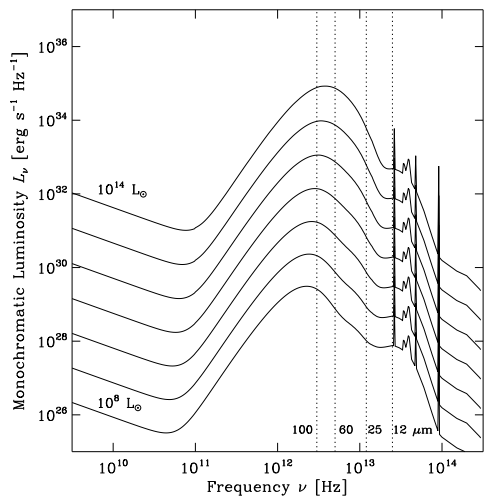}}
\caption
{Fig.\ 1.\ ---
  The assumed galaxy spectral energy distribution in the near infrared to 
  radio wavelengths, constructed from the color--luminosity
  evolution in {\sl IRAS}~galaxies and the tight correlation of the
  far-infrared (FIR) and radio fluxes.
  The prominent emission bands are PAH features.
  Vertical dotted lines represent the wavelengths of {\sl IRAS}~four 
  bandpasses.
  The SEDs with the FIR luminosity of $10^8$, $10^9$, $10^{10}$, $10^{11}$, 
  $10^{12}$, $10^{13}$, and $10^{14}\;L_\odot$ are shown from the 
  bottom in this order.
}
\end{figure}

Now next generation space infrared facilities (e.g. ASTRO-F, {\sl SIRTF}, 
{\sl NGST}~and {\sl FIRST}) are scheduled, and we can expect a vast amount 
of novel observational knowledge with an unprecedented precision.
Among the above facilities, ASTRO-F (Infrared Imaging Surveyor: 
{\sl IRIS}) is the Japanese infrared satellite which will be launched in 2004.
Detailed informations of the ASTRO-F mission are available in 
http://www.ir.astro.isas.ac.jp/ASTRO-F/index-e.html.
In order to calculate the expected detection number in the survey 
planned in the ASTRO-F project, we have made a simple empirical model for 
the galaxy number count (Takeuchi et~al.~1999; Hirashita et~al.~1999).
The applied model was based on the {\sl IRAS}~ surveys.
Therefore, though our previous model has successfully worked,
it cannot reproduce the observed slopes of the novel {\sl ISO}~source 
counts, and the evolutionary effect is no longer satisfactorily estimated.
Thus, we need an improved model to study the detailed 
observational plans and follow-up strategies in other wavelengths.

There are a number of attempts to predict the IR--sub-mm source counts or 
CIRB properties.
Modeling methods of the source counts and the CIRB roughly fall into two
categories:
One is the backward approach, which is based on the local far-IR
(FIR) luminosity function (LF) and the observed spectral energy distribution 
(SED) of galaxies in the FIR -- sub-mm, 
with the assumptions of simple functional forms for the evolution
(e.g.\ Beichman, Helou 1991; Pearson, Rowan-Robinson 1996; 
Malkan, Stecker 1998; Xu et al.~1998; Takeuchi et al~1999).
The other is the forward approach, which is based on the models constructed 
by detailed processes related to the evolution of galaxies with a number
of parameters (e.g.\ Franceschini et al.~1994; Guiderdoni et al.~1998).
Tan, Silk, Balland (1999) presented an interesting approach to combine 
the two methods.
These two different approaches have their own merits and demerits, and 
thus they must interplay with each other.

In this paper, we apply the backward empirical approach to derive the 
evolution of FIR properties of galaxies, but assume no specific functional 
form for the evolution history.
Instead, we treat the galaxy evolution as a stepwise nonparametric form
and search for the most plausible solutions, which well reproduce 
both of the the available IR galaxy number counts and the remarkable
spectrum of the CIRB in a consistent manner.

\begin{figure}[t]
\epsfxsize=7.0cm
\centerline{\epsfbox{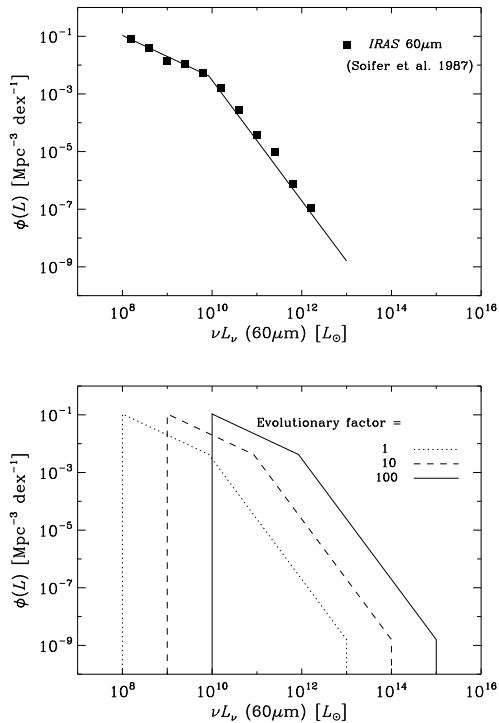}}
\caption
{Fig.\ 2.\ ---
  Upper panel: the applied 60-$\mu$m luminosity function (LF) in this paper. 
  This LF is derived by Soifer, Houck, Neugebauer (1987).
  Lower panel: a schematic representation of its evolution as a function of 
  redshift.
}
\end{figure}

The rest of this paper is organized as follows: We give the model 
description and formulation in Section~2.
We illustrate the method of our analysis in Section~3.
In Section~4 we show the most probable solutions to explain the observed
IR number counts and CIRB.
Some discussions are presented in Section~5.
Section~6 is devoted to our conclusions.
We present the expected number counts of the ASTRO-F FIR all-sky survey
in the Appendix.
Throughout this paper we use the following cosmological parameter set, 
unless otherwise stated: 
$H_0 = 75 \; {\rm km\,s^{-1}\, Mpc^{-1}}$, $q_0 = 0.1$, and $\lambda_0 = 0$.

\section{Model Description}

The galaxy number count model is represented by SED, LF, cosmology, and 
galaxy evolution.
We review the framework of the number count model and discuss these 
ingredients for our model in this section.

The infrared -- radio galaxy SED is modeled by the superposition of 
the following components.
For the infrared -- sub-mm component, we consider PAH (polycyclic aromatic 
hydrocarbon), graphite and silicate dust spectra (Dwek et al.~1997).
In the case of the radio-quiet sources, the emission at millimeter wavelength
regime is dominated by the synchrotron radiation explained by supernova 
remnants (Condon 1992).

By considering the above, we start to construct the infrared model 
SEDs based on the {\sl IRAS}~color--luminosity 
relation (Smith et al. 1987; Soifer, Neugebauer 1991).
Smith et~al.~(1987) and Soifer, Neugebauer (1991) found that 
the {\sl IRAS}~FIR colors and FIR luminosity are tightly correlated.
The FIR color--60-$\mu$m luminosity relation derived by 
Smith et al.~(1987) is as follows:
\begin{eqnarray}
  \log {\displaystyle\frac{S_{60}}{S_{100}}} = (0.10 \pm 0.02) \log L_{60} 
  - (1.3 \pm 0.2)\;,
\end{eqnarray}
where $S_\lambda$ is the detected flux density at wavelength 
$\lambda$ [$\mu$m], and $L_{60}$ [$L_\odot$] is the intrinsic luminosity
evaluated at the $60$-$\mu$m bandpass, 
$L_{60} \equiv \nu L_\nu \;{\rm at}\; 60\;\mu{\rm m}$.
The relation reported by Soifer, Neugebauer (1991) is slightly nonlinear, 
but also a monotonic function of $L_{60}$.
We interpreted this relation to the dust temperature $T_{\rm dust}$--$L_{60}$ 
relation and calculated the modified blackbody continuum with the 
corresponding $T_{\rm dust}$.
We adopted the dust emissivity $\varepsilon_\nu \propto \nu^\gamma$ with 
$\gamma = 1.0, 1.5$, and 2.0, and found the three values yield only a small 
difference ($\ltsim 10$\%) in the $T_{\rm dust}\mbox{--}L_{60}$ relation.
We hereinafter use $\gamma = 1.5$.

We then utilized the data provided by Soifer, Neugebauer (1991) and
derived the approximately linear relation between {\sl IRAS}~flux densities, 
$S_{60}$ and $S_{25}$.
The tight correlation between the FIR and mid-IR (MIR) flux is also found
in the diffuse emission from the Galaxy by Infrared Telescope in Space 
({\sl IRTS}$\,$: Shibai, Okumura, Onaka 2000).
We thus added the mid-infrared spectra proposed by Dwek et~al.\ (1997)
to the FIR component so that the superposed spectra reproduce the correlation
reported by Smith et~al.~(1987), such that $\log S_{25}/S_{60} \simeq -0.9$.
The unidentified infrared bands (UIBs), which we assumed to be produced
by PAHs, are also an important component of the IR SEDs of galaxies.
We set the continuum-to-band intensity ratio as reported by Dwek 
et al.~(1997).
Detailed properties of PAHs are taken from Allamandola et al.~(1989),
e.g.\ we set the PAH features at $3.3\;\mu$m, $6.2\;\mu$m, $7.7\;\mu$m 
with a broader component at $5.5 - 9.5\;\mu$m, and $11.3\;\mu$m.

A remarkably tight and ubiquitous correlation is well-known between 
the FIR continuum flux and radio continuum flux (e.g.\ Helou, Soifer, 
Rowan-Robinson 1985; Bregman, Hogg, Roberts 1992; Condon 1992).
For the longer wavelength regime, a power-law continuum produced by 
synchrotron radiation ($\propto \nu^{-\alpha}$) dominates the observed 
emission.
We set $\alpha = 0.7$ according to Condon (1992), and added the FIR composite
spectra, using the relation
\begin{eqnarray*}
  q \hspace{-3mm}&\equiv& \hspace{-3mm}
  \log \left( \frac{\mbox{FIR}}{3.75 \times 10^{12}\;\mbox{W m}^{-2}}\right)
  -\log \left( \frac{S_{1.4\,{\rm GHz}}}{{\rm Wm^{-2}Hz^{-1}}} \right)\\
  &=& \hspace{-3mm}2.3\;, 
\end{eqnarray*}
where
\begin{eqnarray}
  \mbox{FIR} \; [{\rm Wm^{-2}}] &\equiv& 1.26 \times 10^{-14}\nonumber\\
  &&\times\left( 2.58 S_{60} \; [\mbox{Jy}] + S_{100} \;[\mbox{Jy}] \right)
    \label{eq:helou1}
\end{eqnarray}
(Condon 1992).
Equation~(\ref{eq:helou1}) is adopted in Helou et al.~(1988) and
will be applied again in Section 5.
This is the final SED we use in our number count and CIRB models 
(see Figure~1).

\subsection{Local Luminosity Function and Evolutionary Effect}

We adopted the $60$-$\mu$m LF based on the {\sl IRAS}~ by Soifer et al.~(1987)
as the local IR LF of galaxies:
\begin{eqnarray}
  &&\hspace{-1cm}\log \phi_0(L_{60}\;[L_\odot]) =\nonumber \\
  \hspace{-3mm}&&\hspace{-3mm}\left\{
    \begin{array}{l}
      4.87 - 0.73 \log L_{60} \,[L_\odot] \\
      \hspace{2cm}\mbox{for}\;10^8 < L_{60} < 10^{9.927}; \\
      18.5 - 2.1 \log L_{60} \,[L_\odot] \\
      \hspace{2cm}\mbox{for}\;10^{9.927} < L_{60} < 10^{13}; \\
      {\rm no \; galaxies} \\
      \hspace{2cm}\mbox{otherwise},
    \end{array}
  \right.\label{lf}
\end{eqnarray}
where $\phi_0$ is the number density of galaxies in Mpc$^{-3}$dex$^{-1}$.
We show the LF in the upper panel of Figure~2.
We applied the double power-law form for the local LF.
The lower cut-off luminosity do not seriously affect the result, as stated 
in Takeuchi et al.~(1999).
The effect of the faint-end slope of the LF on the number count estimation
is briefly summarized in Takeuchi, Shibai, Ishii (2000a).

We assumed a pure luminosity evolution in this study.
In this case 60-$\mu$m luminosity of a certain galaxy at redshift $z$
is described as
\begin{eqnarray}
  L_{60} (z) = L_{60} (0) f(z)\;, \label{deflev}
\end{eqnarray}
We also assumed that the luminosity evolution is `universal', i.e.\ 
independent of galaxy luminosity.
This is depicted in the lower panel of Figure~2.
In the previous backward approach, a certain functional form 
$f(z)$ was often assumed for the evolution of galaxies, and model counts 
are calculated and compared with observations.
We, on the contrary, treat the evolution of galaxy luminosities 
as a stepwise nonparametric form which is to be statistically estimated, 
in order to explore the most suitable evolutionary history which 
reproduces the present observational results.
Mathematically, we regard the evolution estimation from the number count 
as an inverse problem, and we apply the universal LF evolution in order 
to obtain a statistically stable solution.
Constraints of $f(z)$ from the observed CIRB spectrum will be given in 
Section~4.1.
Another approach mainly based on the density evolution is discussed in
Pearson et~al.~(2000).
We note that the present quality of the IR number count data is not 
sufficient to estimate the evolution of the shape of the LF, and fine tuning
of the LF shape is less meaningful.
Such works remain to be solved after the forthcoming release of huge 
databases.

\subsection{Formulation of the Galaxy Number Count}

Using the above formulae, now we calculate the flux--number 
($\log N\mbox{--}\log S$) relation, or the so-called 
galaxy number count.
We assume that galaxies can be regarded as point sources (i.e.\ the 
cosmological dimming of the galaxy surface brightness is not taken into 
account).
Then the relation between the observed flux $S(\nu)$ and the emitted 
monochromatic luminosity $L(\nu_{\rm em}) = L((1 + z)\nu )$ is given by 
\begin{eqnarray}
  S(\nu ) =\frac{(1+z)L\left( (1+z)\nu\right)}{4\pi d_{\rm L}^2}\,,
  \label{detec}
\end{eqnarray}
where $d_{\rm L}$ is the luminosity distance.
When we fix a certain $S(\nu)$, we obtain $L((1 + z) \nu)$ by using 
equation~(\ref{detec}).
Then the corresponding $L_{60}(S(\nu),\; z)$ at a redshift $z$ is 
uniquely determined.
We define $N(> S (\nu))$ as the number of galaxies with a detected 
flux density larger than $S (\nu)$ in a survey solid angle $\Omega$, 
then it is formulated as
\begin{eqnarray}
  &&\hspace{-1.2cm}N(>S(\nu))=\nonumber \\
  &&\hspace{-1.0cm}\int_\Omega {\rm d}\Omega\int_0^{z_{\rm max}}{\rm d}z
  \frac{{\rm d}^2V}{{\rm d}z\,{\rm d}\Omega}
  \int_{L_{60} (S(\nu)\,, z)}^{\infty}
  \phi(z,\, L_{60}') \; {\rm d} L_{60}'\; , \label{gnc}
\end{eqnarray}
where ${{\rm d}^2V}/{{\rm d}z}{\rm d}\Omega$ is the comoving volume 
element per sr, and $z_{\rm max}$ is the maximum redshift we consider in this
study.
As we see later, we set $z_{\rm max} = 5$.

We then formulate the number count with an evolution.
If a pure luminosity evolution takes place, by using equation~(\ref{deflev}),
the evolution of the luminosity function with redshift is expressed as
\begin{eqnarray}
  \phi (z,L_{60} )\, {\rm d} L_{60} = 
  \phi_0 \left( \frac{L_{60}}{f(z)} \right) 
  \;{\rm d} \left( \frac{L_{60}}{f(z)} \right)
\end{eqnarray}
where $\phi_0(L_{60})$ is the IR LF in the local Universe.
The expected number count is expressed as 
\begin{eqnarray}
  &&\hspace{-1.2cm}N(> S (\nu) ) \nonumber \\
  &&\hspace{-1.0cm}=\int_\Omega {\rm d} \Omega \int_0^{z_{\rm max}}{\rm d}z
  \frac{{\rm d}^2V}{{\rm d}z\,{\rm d}\Omega}
  \int_{L_{60}(S(\nu), z)}^{\infty} \;\hspace{-5mm}
  \phi_0 \left( \frac{{L_{60}}'}{f(z)} \right)\; 
  {\rm d} \left( \frac{{L_{60}}'}{f(z)} \right) \nonumber \\
  &&\hspace{-1.0cm}=\int_\Omega {\rm d} \Omega \int_0^{z_{\rm max}}{\rm d}z
  \frac{{\rm d}^2V}{{\rm d}z\,{\rm d}\Omega}
  \int_{L_{60}(S(\nu), z)/f(z)}^{\infty} \;\hspace{-5mm}
  \phi_0 (\tilde{L}_{60}')\; {\rm d} \tilde{L}_{60}'
\end{eqnarray}
where $\tilde{L}_{60} = L_{60} / f(z)$.
The full formulation of the galaxy number count including a density evolution 
is given in e.g.\ Gardner (1998) or Takeuchi et al.\ (1999).
In this paper we call $f(z)$ `the evolutionary factor'.

\subsection{Formulation of the Cosmic Infrared Background}

The CIRB is generated from the integrated light of galaxies.
Therefore, combining the SEDs of galaxies and number count predictions, 
we obtain the CIRB spectrum.
The observed flux density of a galaxy whose IR luminosity is $L_{60}$, 
$S(\nu, L_{60})$, is given by equation (\ref{detec}) as follows:
\begin{eqnarray}
  S(\nu \,,L_{60})=\frac{(1 + z)L\left(\nu (1+z)\,, L_{60} \right) }
  {4\pi d_{\rm L}^2},
\end{eqnarray}
where $L(\nu , L_{60})$ is the monochromatic luminosity of a galaxy with 
$L_{60}$.
Then the CIRB spectrum $I(\nu)$, i.e., the background flux density from 
a unit solid angle, is expressed as
\begin{eqnarray}
  \hspace{-7mm}I(\nu )=\int_0^{z_{\rm max}} {\rm d}z 
  \frac{{\rm d}^2V}{{\rm d}z\,{\rm d}\Omega}
  \int_0^\infty\phi (z,\, L_{60}') S(\nu\,, \;L_{60}') \; 
  {\rm d} L_{60}' \;.
\end{eqnarray}
We can deal with the evolutionary effect on $I(\nu)$ through 
$\phi(z, L_{60})$, just the same as in the case of the number count 
calculation.

\section{Analysis}

In principle the galaxy number count and the spectral shape of the CIRB are
integrated values along with redshift.
Therefore, the information of galaxy evolution with redshift is degenerate
if we analyze a dataset obtained from a single wavelength.
But the redshift degeneracy can be solved by treating the multiband
observational results at the same time as suggested by 
Takeuchi et~al.~(2000b), who studied the changes of the IR number counts 
against some extreme galaxy evolution histories.

Here we thoroughly survey the response of the multiband number counts and 
the CIRB to the change of the evolutionary factor at various redshifts.
First we divide the redshift into the follwoing eight intervals, [0 -- 0.25], 
[0.25 -- 0.5], [0.5 -- 0.75], [0.75 -- 1.0], [1.0 -- 1.5], [1.5 -- 2.0], 
[2.0 -- 3.0], and [3.0 -- 5.0].
If we set the evolutionary factor $f(z) = 10$ at each interval, 
the number counts and the CIRB will deviate from the no-evolution prediction
differently, corresponding to each interval.
This different response of the predictions provides a useful information
of the evolution with redshift.
For example, consider a certain deviation in the number count at one 
waveband.
Without any other informations, we cannot distinguish a certain evolution 
history from all the other possibilities.
But when we use the deviations in other wavelengths and in the CIRB,
we can check the validity of an evolution by the goodness of fit to the data
at other wavelengths and may infer a much more constrained parameter range.

We show the deviations of the CIRB spectrum and the galaxy number counts 
with an order-of-magnitude evolution at each redshift interval in 
Figures~3 and 4.
The left panel of Figure~3 depicts the response of the CIRB spectrum
for hypothetical, an order-of-magnitude evolution at the above redshift
interval at $z < 1$.
The no-evolution prediction of the CIRB is presented by the thick solid 
curve.
The dotted line represents the CIRB when $f(z) = 10$ at $0 < z < 0.25$, 
the dashed line at $0.25 < z < 0.5$, the dot-dashed line at $0.5 < z < 0.75$,
and the dot-dot-dot-dashed line at $0.75 < z < 1.0$, respectively, in 
Figure~3.
The peak of each CIRB model prediction shifts toward longer wavelengths,
with the increasing redshift of the input test evolution, while
the intensity of the CIRB at sub-mm regime is insensitive to the evolution
at $z < 1$.
On the other hand, the right panel of Figure~3 illustrates the response
of the CIRB caused by each test evolution at $z > 1$.
This time the dotted line represents the CIRB when $f(z) = 10$ at 
$1.0 < z < 1.5$, the dashed line at $1.5 < z < 2.0$, the dot-dashed line 
at $2.0 < z < 3.0$, and the dot-dot-dot-dashed line at $3.0 < z < 5.0$, 
respectively.
Again the thick solid line is the no-evolution prediction.
At $z > 1$, the evolution mainly affects the intensity of the sub-mm CIRB
spectrum.

\begin{figure*}[t]
\epsfxsize=12.0cm
\centerline{\epsfbox{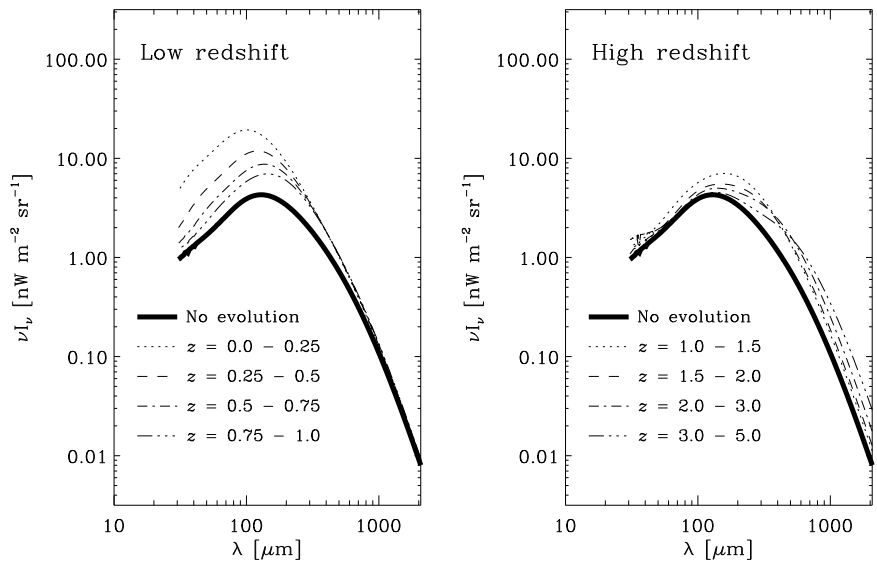}}
\caption
{Fig.\ 3.\ ---
  The deviations of the CIRB spectrum and the galaxy number counts 
  with an order-of-magnitude evolution at the following eight intervals: 
  [0 -- 0.25], [0.25 -- 0.5], [0.5 -- 0.75], [0.75 -- 1.0], [1.0 -- 1.5], 
  [1.5 -- 2.0], [2.0 -- 3.0], and [3.0 -- 5.0].
  The left panel depicts the response of the CIRB spectrum
  for hypothetical, an order-of-magnitude evolution at the above redshift
  interval at $z < 1$.
  The no-evolution prediction of the CIRB is presented by the thick solid 
  curve.
  The right panel illustrates the response caused by each test evolution at 
  $z > 1$.
}
\end{figure*}

\begin{figure*}[t]
\epsfxsize=14.0cm
\centerline{\epsfbox{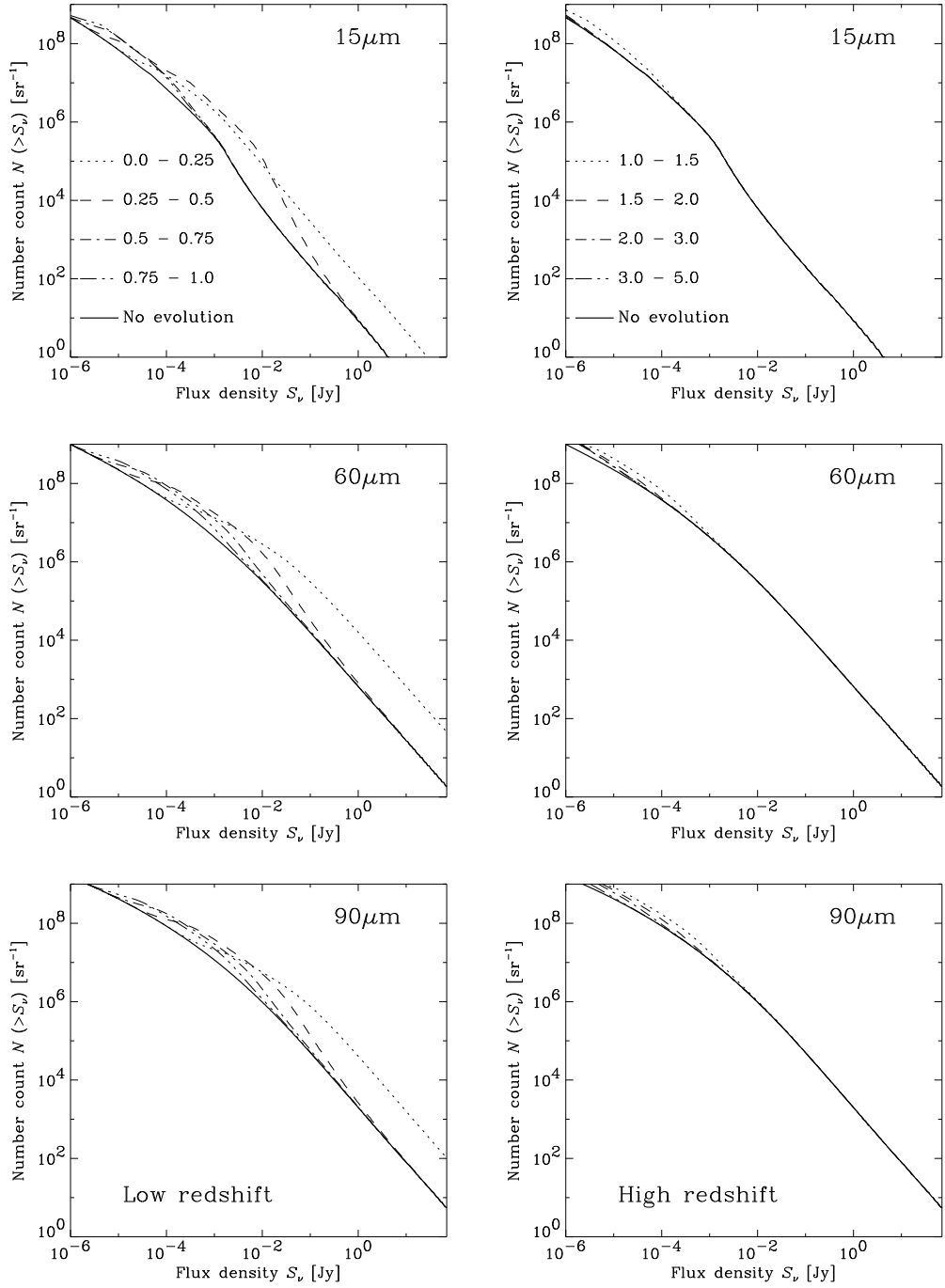}}
\caption
{Fig.\ 4a.\ ---
  The effect of the test evolution to the multiband number counts.
  The response galaxy number counts at $15\;\mu$m, $60\;\mu$m, and $90\;
  \mu$m are displayed.
  The same as in Figure~3, the response of the number counts for the input
  test evolution at $z < 1$ are indicated in the left panels, and that for 
  the input evolutions at $z > 1$ in the right panels.
}
\end{figure*}

\begin{figure*}[t]
\epsfxsize=14.0cm
\centerline{\epsfbox{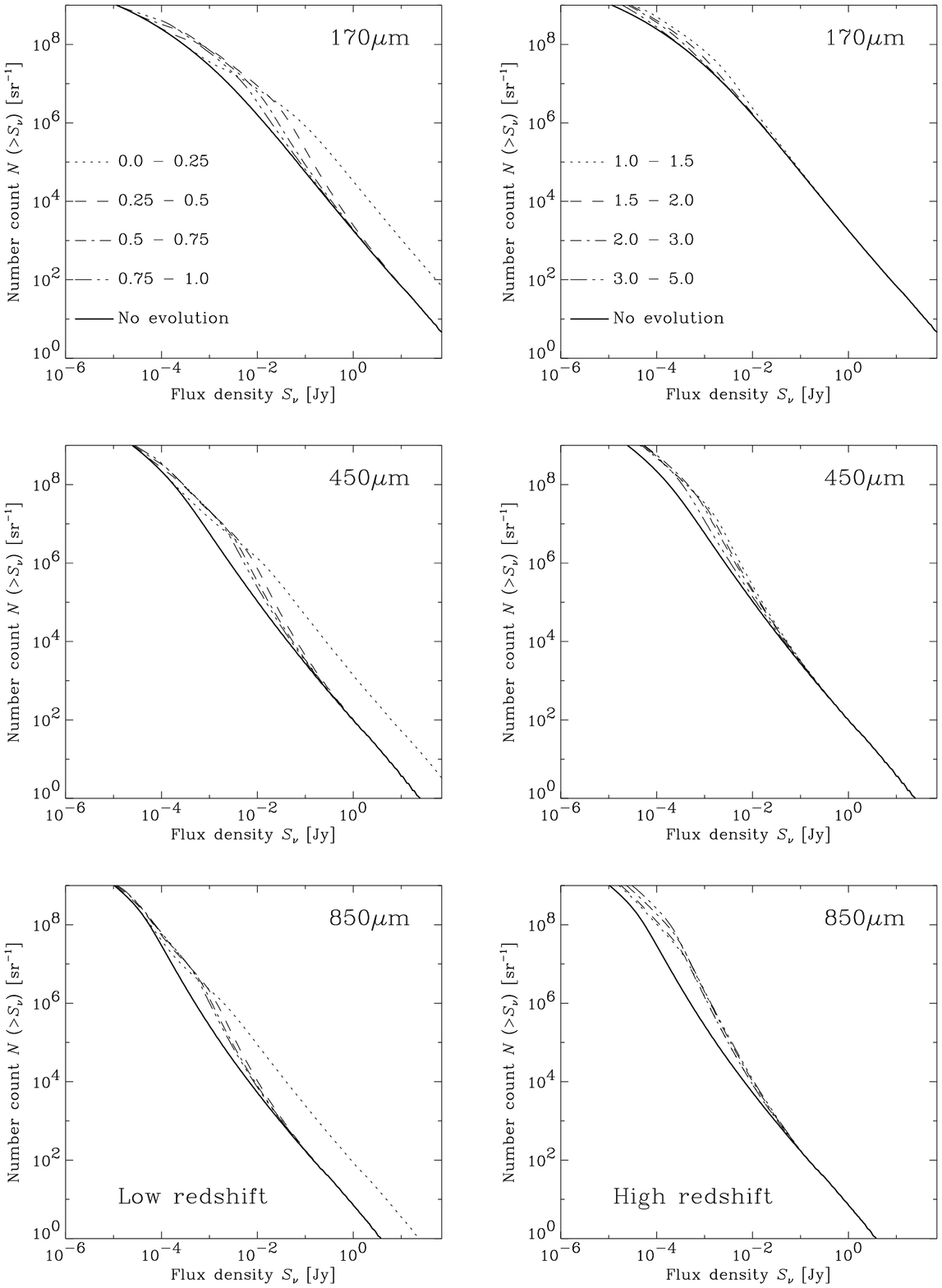}}
\caption
{Fig.\ 4b.\ ---
  The same as Figure~4a, except that these counts are at $170\;\mu$m, 
  $450\;\mu$, and $850\;\mu$m.
}
\end{figure*}

How about the number count?
We show the effect of the test evolution to the multiband number counts
in Figure~4.
The response galaxy number counts at $15\;\mu$m, $60\;\mu$m, and $90\;\mu$m 
are displayed in Figure~4a, while those at $170\;\mu$m, $450\;\mu$, and
$850\;\mu$m are in Figure~4b.
The same as in Figure~3, the response of the number counts for the input
test evolution at $z < 1$ are indicated in the left panels, and that for 
the input evolutions at $z > 1$ in the right panels, in Figure~4a and 4b.

We clearly see that the MIR 15-$\mu$m count is almost independent of 
the status of galaxies at $z \gtsim 0.75$, and it is quite sensitive to 
the evolution at $0.25 < z < 0.5$.
Thus the 15-$\mu$m galaxy number count is a useful and unique indicator
of the evolution at such low redshifts.
The 60-$\mu$m count depends still strongly on the evolution at 
$0.25 < z < 0.5$, and is more sensitive to the evolutionary status at
$0.5 < z < 0.75$.
On the contrary it is insensitive to that at $z \gtsim 1$.
This trend of the redshift dependence becomes more prominent at FIR
90- and 170-$\mu$m number counts.
The effect of the evolution at $1.0 < z < 1.5$ begins to appear in
the faintest slope of the number counts at these wavelengths.
The sub-mm 450- and 850-$\mu$m counts shows, in contrast, a weaker
dependence on the low-$z$ ($z < 1$) galaxy evolution.
They are very strongly dependent on the evolutionary factor at 
$z \gtsim 1$.
Here, notice that the effect of the test evolution at $z > 1$ 
are almost constant in 850 $\mu$m.
This is caused by the counter-intuitive, negative $K$-correction.

In summary, we are able to resolve the degeneracy of the galaxy evolution 
history by treating the multiband datasets and the CIRB at the same time.
Since the amplitude of the evolutionary factor at various redshift intervals
correlates with each other, this statistical inference problem is still more 
difficult than a mere fitting problem.
But we can obtain a set of permitted ranges of the evolutionary factors 
as a function of redshift by this method.

\section{Results}

In Figure~5, we summarize the observational constraints of the evolutionary 
factor estimated from the IR galaxy number counts and the CIRB.
Thick dashed horizontal lines depict the bounds mainly based on
the IR galaxy number counts.
Thick solid lines are the constraints based on both CIRB and IR number counts.
Thick dot-dashed line is determined by the severe constraints from CIRB
spectrum at sub-mm wavelengths.
We set that there are no galaxies at $z > 5$.
This is not a strict observational constraints, but the observed CIRB 
suggests that there are few luminous IR galaxies at such a high redshift.
We explain and discuss the details of these constraints in the following.

\subsection{Constraint from Cosmic Infrared Background}

Today we have a considerable information on the spectrum of the CIRB.
As we mentioned in Section~1, the CIRB has been revealed to have a 
surprisingly high intensity at FIR wavelength regime 
($\nu L_\nu = 25 \pm 7$ nW$\, {\rm m^{-2}sr^{-1}}$ at 140 
$\mu$m and $\nu L_\nu = 14 \pm 3$ nW$\,
{\rm m^{-2}sr^{-1}}$ at 240 $\mu$m: Hauser et al.~1998).
Thus the spectral shape of the CIRB can be used to constrain the evolutionary 
history of the cosmic IR luminosity density which is closely related to the 
cosmic history of star formation and metal production.

We summarize the reported intensity of the CIRB in Figure~6.
Open squares with downward arrows represent the ``dark sky'' upper limit 
to the CIRB measured by {\sl COBE}~Diffuse Infrared Background Experiment 
(DIRBE).
The DIRBE~sky brightness varies roughly sinusoidally over the year, 
due to the complex features of the interplanetary dust cloud.
The ``dark sky'' brightness is that of the darkest area on the sky at each
wavelength.
Open squares with upward arrows represent the residual signal of DIRBE after 
removing the contributions from the model foreground sources such as Galactic 
diffuse emission, and interplanetary dust emission.
Filled squares with error bars are the DIRBE detections at 140 $\mu$m and 
240 $\mu$m by Hauser et al.~(1998).
The hatched region is the permissible range derived from the {\sl COBE}~Far 
Infrared Absolute Spectrophotometer (FIRAS) high-frequency data, 
after removal of the cosmic microwave background (CMB) signal 
(Fixsen et al.\ 1998).
At wavelength shorter than 140 $\mu$m, the strong emission from the 
interplanetary dust prevents us from estimating the reliable value of the 
extragalactic component, and therefore in this study, we do not use the 
intensity at this wavelengths recently reported by some authors.
The detector noise level of DIRBE~becomes worse at wavelength longer than 
240 $\mu$m.
On the other hand, the CIRB spectrum at sub-mm regime detected by FIRAS~is 
considered to be highly robust, because in this wavelength regime, what 
to be subtracted is only the blackbody spectrum of the CMB, which has no model
ambiguity.

\begin{figure}[t]
\epsfxsize=7.0cm
\centerline{\epsfbox{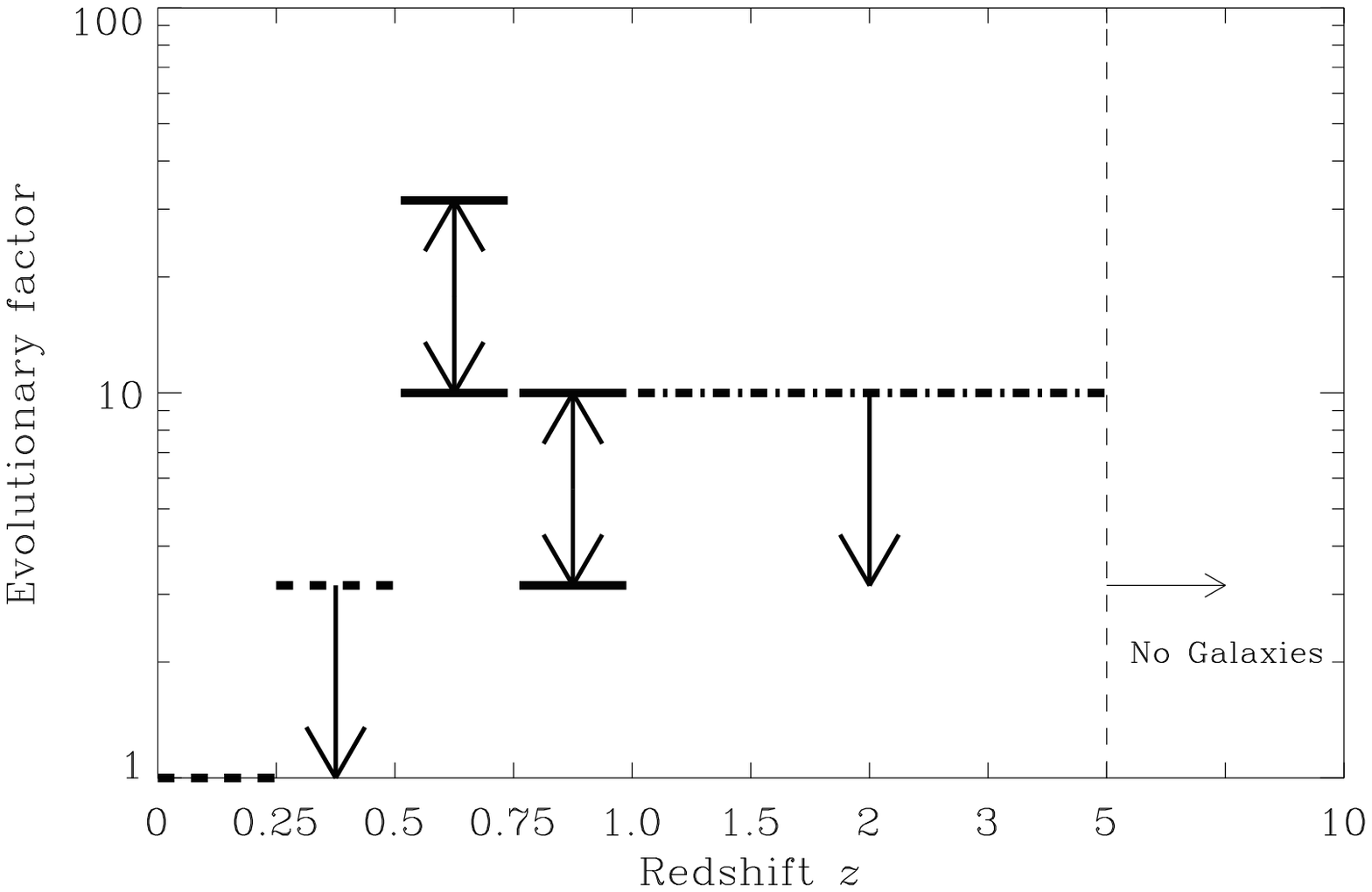}}
\caption
{Fig.\ 5.\ ---
  The summary of the constraints of the galaxy luminosity evolution
  inferred from the infrared galaxy number counts and the cosmic infrared 
  background radiation spectrum.
}
\end{figure}

Lagache et al.~(1999) claim that, from their observation of the sky in and 
around the Lockman Hole, there exists a significant contribution
from the Galactic warm interstellar medium (WIM) to the IR background 
intensity.
Thus if we adopt their interpretation, the CIRB intensities become
$15.3 \pm 6.4 \; {\rm nW\,m^{-2}sr^{-1}}$ 
(140 $\mu$m) and $11.4 \pm 1.9\; {\rm nW\,m^{-2}sr^{-1}}$
(240 $\mu$m).
The contribution of the WIM at sub-mm wavelength is small, and the intensity
of the CIRB at $200 - 2000 \; \mu$m is almost unchanged from Fixsen et al.'s
value.
However, in other regions, Lagache et al.~(2000) reported similar results to 
Hauser et al.~(1998) even though they again consider the contributions of 
the WIM, and the existence of the WIM still seems a matter of debate.
We present the claimed value of Lagache et al.~(1999) by filled triangles
with errors in Figure~6.

At a glance of Figure~6, it is obvious that no evolution prediction is 
strongly ruled out.
The CIRB intensity at $140\;\mu$m suggests an order-of-magnitude evolution
of IR galaxies.
It is important that such a short wavelength of the CIRB peak can only be 
reproduced by a rapid evolution of galaxies at low redshift ($z < 1$).
Even if we use the slightly lower value of Lagache et al.~(1999), 
this conclusion is not affected.
If we try to explain the peak intensity by high-redshift IR galaxies, 
vast numbers of galaxies is required.
Furthermore, if in case, the CIRB intensity at longer wavelength have to be 
overpredicted, and the observed spectrum will be seriously violated.
As we noted that the FIRAS spectrum at sub-mm wavelength is highly 
reliable, all of the model must reproduce it exactly.
At least, models which overestimate FIRAS spectrum should be dismissed.
The sub-mm slope of the CIRB is shallower compared with a single galaxy SED,
which should be interpreted as an integrated SED of a significant numbers 
of galaxies at high redshift ($z > 2-3$) (see Figure~3).
Thus, though the constraint is strong, it is not inconsistent with the 
actual existence of obscured high-$z$ starbursts (e.g.\ Hughes~2000; 
Ishii et al.~2000).
What we should stress is that there is an observational upper bound
of the evolution of IR galaxies.
In Figure~5, the constraints from the CIRB peak intensity is presented as 
the enhance of the evolutionary factor at $0.5 < z < 0.75$, 
and the constraints from FIRAS~observation is shown by the upper bound
at $1 < z < 5$.

Hereafter, within the permitted range derived from the CIRB, we focus on 
the three representative evolutionary histories.
We show these evolutions in Figure~7.
Evolution~1 has a steep rise at $0 < z < 0.5$, reaches the summit at which
the evolutionary factor $f(z)$ is 10, and has a plateau at $0.5 < z < 2$.
Then, at $2 < z$, the evolutionary factor slowly decreases but still
higher than the local value.
This rather moderate evolution is consistent with the CIRB $140$-$\mu$m
intensity when we consider the WIM contribution, and slightly overestimates
the FIRAS sub-mm spectrum.
These features are also described in Figure~6 as the dot-dashed line.

\begin{figure}[t]
\epsfxsize=7.0cm
\centerline{\epsfbox{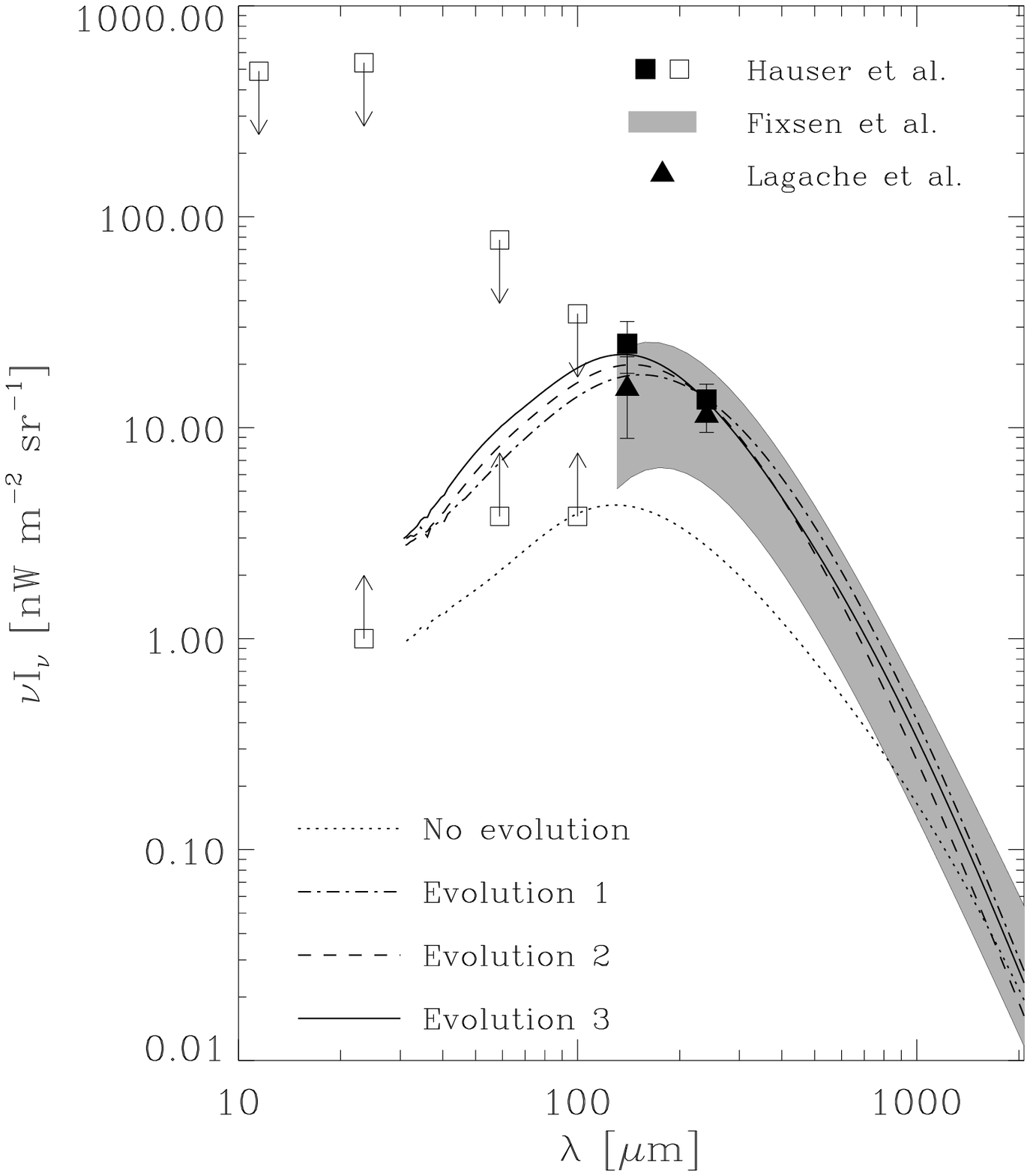}}
\caption
{Fig.\ 6.\ ---
  The cosmic infrared background observations and our model predictions.    
  Dotted line describes the no-evolution prediction.
  Dot-dashed line shows the model CIRB with Evolution~1, dashed line
  represents the CIRB with Evolution~2, and solid line, with Evolution~3, 
  respectively.
}
\end{figure}

Evolution~2 has a stronger rise of the evolutionary factor than Evolution~1
at $0 < z < 0.5$.
Beyond the peak at $0.5 < z < 0.75$, the evolutionary factor decreases 
rapidly to the local value at $z < 2$, and at $2 < z < 5$ the IR LF is the 
same as the local one.
This evolution mitigates the overestimation of the sub-mm CIRB, and 
the $140$-$\mu$m peak intensity has an intermediate value between Hauser et
al.~(1998) and Lagache et al.~(1999).

Finally Evolution~3 has the strongest rise at $0 < z < 0.5$.
This evolution model has a prominent peak at $0.5 < z < 0.75$, which well 
reproduces the $140$-$\mu$m peak of the CIRB by Hauser et al.~(1998).
This evolution also has a long plateau at $0.75 < z < 5$.
The predicted CIRB by this model is quite consistent with the FIRAS~spectrum.
Since we present the evolutionary history as a function of redshift, we
would have an impression that the rise of the evolutionary factor at 
$0.5 < z < 0.75$ is unnaturally steep, but the peak height is 30
and the duration is $\sim 2$ Gyr, both of which are within the bounds 
of possibility and worth being considered.

These CIRB model behavior is also shown in Figure~6.
No-evolution prediction is represented by dotted line in Figure~6.
The dot-dashed line is the model CIRB produced by Evolution~1, 
the dashed line is produced by Evolution~2, and the solid line, 
by Evolution~3.

\subsection{Constraint from IR Galaxy Number Counts}

We examine whether the three evolution models are consistent with
observed galaxy number counts.
Recently, new observational results of the galaxy number counts have become
available at MIR, FIR, and sub-mm, mainly by the success of {\sl ISO}~(Infrared
Space Observatory) and SCUBA (Submillimeter Common-User Bolometer Array for 
the James Clerk Maxwell Telescope).
We are able to compile these data as well as previously obtained {\sl IRAS} 
databases.

We present the galaxy number counts at $15\;\mu$m -- $850\;\mu$m in Figure~8.
Again the no-evolution predictions are represented by dotted lines, 
the dot-dashed lines are the model number counts produced by Evolution~1, 
the dashed lines are produced by Evolution~2, and the solid line by 
Evolution~3, respectively.
Here we go into the details of the galaxy count at each wavelength in 
IR -- sub-mm.

\subsubsection{15-$\mu$m galaxy number counts}

The top-left panel of Figure~8 shows the 15-$\mu$m galaxy counts predicted 
by our model with Evolutions~1, 2, and 3.
We also show the available observed $15$-$\mu$m ISOCAM galaxy counts 
(Flores et~al.~1999a, b; Clements et~al.~1999; Aussel et~al.~1999; Altieri
et~al.~1999; Elbaz et~al.~1999; Oliver et~al.~2000a) and the constraint 
from fluctuation analysis ($P(D)$ analysis) of the Hubble Deep Field (HDF)
(Oliver et~al.~1997), together with {\sl IRAS}~galaxy count (Rush, Malkan,
Spinoglio 1993).
The observed counts are well reproduced by any of our three evolutionary
histories, and clearly different from the no-evolution prediction.
The slope and normalization of the {\sl IRAS}~count must be reproduced by the 
model {\sl without any corrections}.
The ELAIS (European Large-Area {\sl ISO}~Survey: Oliver et al.\ 2000b; for the 
present status, see http://athena.ph.ic.ac.uk)
count (Serjeant et~al.~2000) severely constrain the range of the evolutionary 
factor, and this shows that the rapid rise at $z < 0.75$ is necessary to 
explain the MIR galaxy number counts.
Of course too large evolutionary factor is also forbidden by this data.
The permitted range of the evolutionary factor is $f(z = 0.5 \mbox{--} 0.75)
 = 30 \pm 3$ if we take both the Poisson fluctuation and the cosmic variance
into account (see Appendix).

\begin{figure}[t]
\epsfxsize=7.0cm
\centerline{\epsfbox{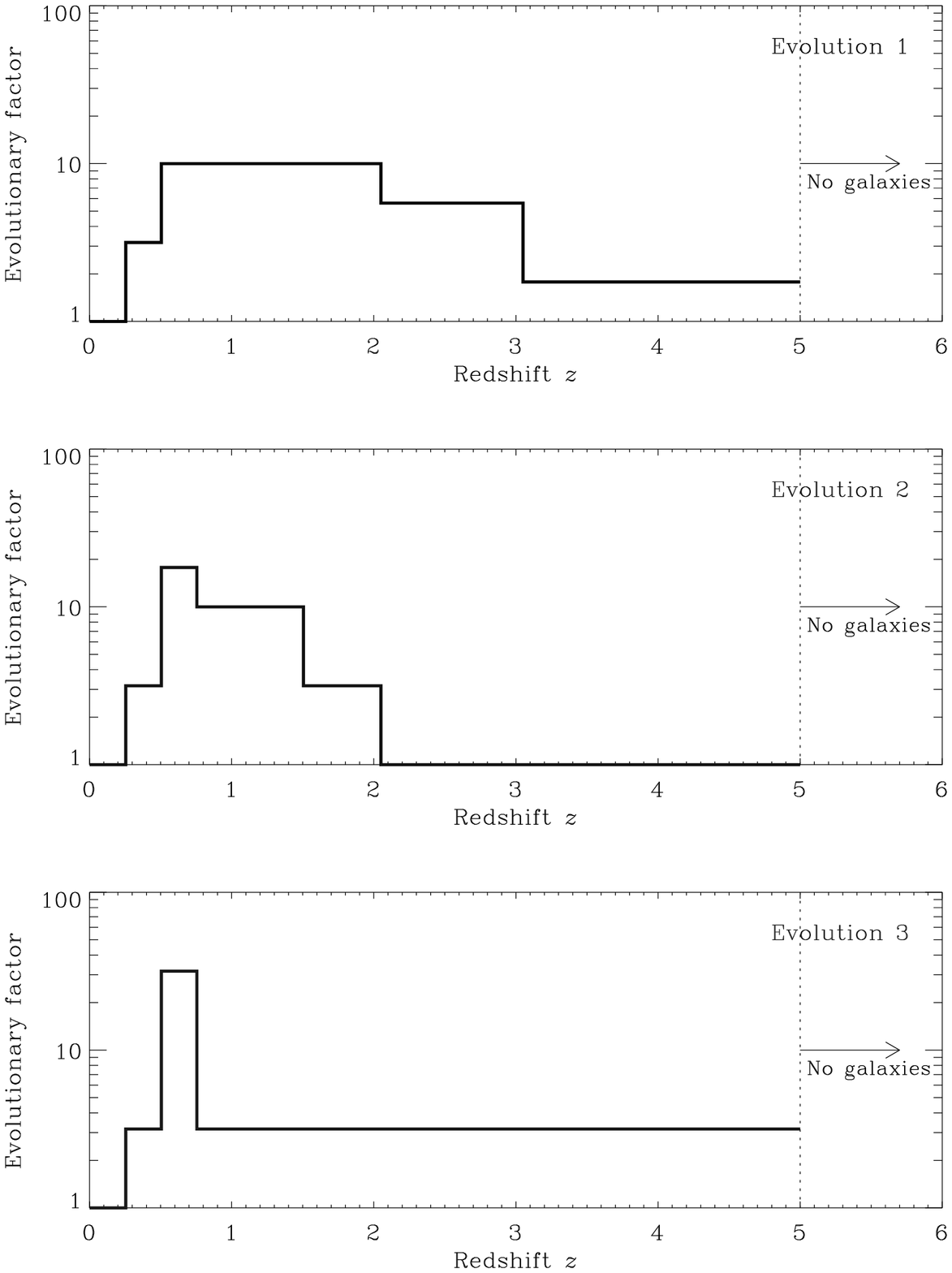}}
\caption
{Fig.\ 7.\ ---
  Three representative galaxy evolutionary histories 
  permitted by the observational constraints presented in Figure~5.
}
\end{figure}

\begin{figure*}[tp]
\epsfxsize=16.0cm
\centerline{\epsfbox{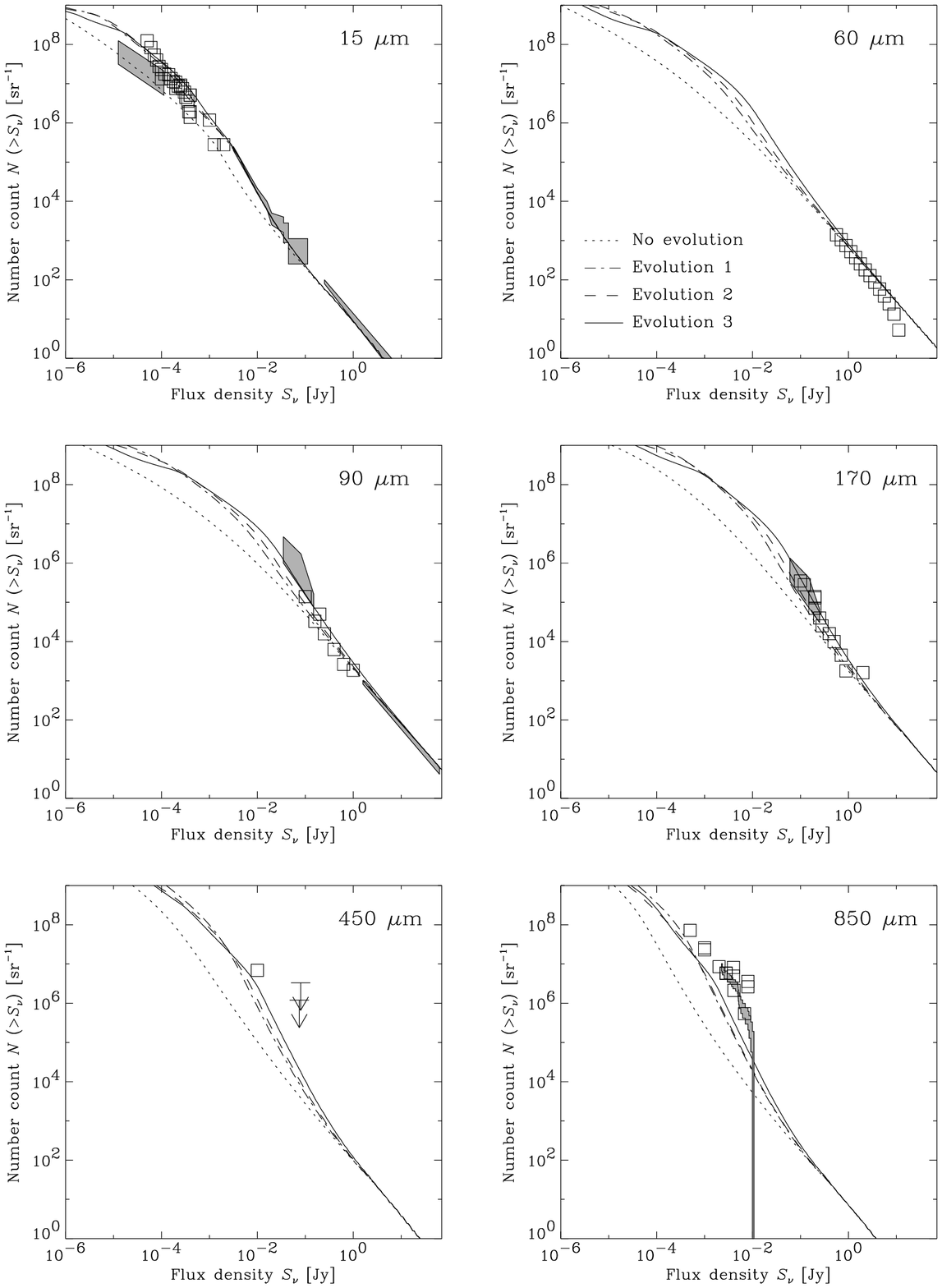}}
\caption
{Fig.\ 8.\ ---
  The multiband galaxy number count at infrared -- sub-mm 
  wavelengths.
  Same as in Figure~6, dotted lines represent the no-evolution prediction.
  Dot-dashed lines are the counts with Evolution~1, dashed lines are those 
  with Evolution~2, and solid lines, with Evolution~3, respectively.
}
\end{figure*}

\subsubsection{Far-infrared galaxy number counts}

Next we examine the galaxy counts at FIR wavelengths.
The top-right panel of Figure~8 is the {\sl IRAS}~60-$\mu$m 
galaxy number count (Rowan-Robinson et~al.~1991).
Our model calculations are based on the statistical studies on the local 
{\sl IRAS}~galaxies and the {\sl IRAS}~60-$\mu$m LF, the 60-$\mu$m count 
is automatically reproduced at the brightest end.
On the other hand, the count at the faint flux limit restricts the evolution
at lowest redshifts ($z\sim 0.1 \mbox{--} 0.3$).
Some previous authors claimed the evidence of the FIR galaxy evolution at
the flux limit of {\sl IRAS} (e.g.\ Saunders et~al.~1990; Ashby et~al.~1996; 
Bertin, Dennefeld, Moshir 1997; Springel, White 1998), but the reported 
strengths of the evolution were not consistent with each other.
We found that the evolutionary factor can be at most $f(z) < 3$ in the 
redshift range $0.25 < z < 0.5$, otherwise the model number count violate
the 1-$\sigma$ variance range of the observed {\sl IRAS} count.

At 90 and 170 $\mu$m, deep galaxy counts are obtained by ISOPHOT.
The photometric calibration of the {\sl ISO}~is still a matter of debate, and
we should take care of the interpretation of the counts at these wavelengths.
The middle-left panel presents our model 90-$\mu$m galaxy counts with
observed FIRBACK source counts (Dole et~al.~2000) and ISOPHOT CIRB project
(Juvela, Mattila, Lemke 2000), and the constraint from the fluctuation 
analysis of the {\sl ISO}~Lockman Hole data (Matsuhara et~al.~2000).
We also show the {\sl IRAS}~100 $\mu$m galaxy counts by a hatched thin area 
at brightest flux.
Again our model shows an excellent agreement with {\sl IRAS}~count.
The middle-right panel is the 170-$\mu$m galaxy counts.
Observed counts are taken from Stickel et~al.~(1998), Kawara et~al.~(1998), 
Puget et~al.~(1999)(FIRBACK), Oliver et al.~(2000)(ELAIS), 
Juvela et al.~(2000) and results of the fluctuation analysis of the 
Lockman Hole (Matsuhara et~al.~2000).
Most of these data suggest that the slope of the number counts is very steep,
i.e.\ ${\rm d} \log N/{\rm d} \log S \ltsim -2.5$.
This steep slope index strongly supports the rapid evolution toward 
$z \sim 0.75$.
This suggested evolution is quite consistent with the requirements of 
the CIRB spectrum.

We note that the Evolutions 1, 2, and 3 show significant difference at 
$0.1 - 0.01$ Jy in the FIR wavelength.
Especially, only Evolution~3 can reproduce the constraint of Matsuhara 
et~al.~(2000).
Thus a huge database, such as the data of ASTRO-F all-sky survey, can 
discriminate this difference and fix the evolutoinary history at this
redshift regime ($0\ltsim z\ltsim 0.75$).

\subsubsection{Submillimeter galaxy number counts}

The bottom-left and bottom-right panels of Figure~8 show the 450- and 
850-$\mu$m model number counts and observed counts obtained by SCUBA.
It is now well accepted that the sub-mm source counts are suitable for 
examine the status of the high-redshift ($z > 1 - 2$) galaxies.
The observed $450\;\mu$m counts are recently obtained by Blain et al.~(2000), 
and two upper limits are based on Smail et~al.~(1997) and Barger, Cowie, 
Sanders (1999a).
The symbols represent the observed $850\;\mu$m counts taken from 
Smail, Ivison, Blain (1997), Hughes et al.~(1998), Blain et~al.~(1999), 
Eales et~al.~(1999) and Holland et~al.~(1999),
and the hatched region, Barger et al.~(1999a).

The 450-$\mu$m counts of Blain et al.~(2000) can be reproduced by 
Evolution~3, because this evolutionary history has a long plateau toward
$z = 5$.
On the other hand, the 850-$\mu$m counts are somewhat problematic.
If we attach importance to the FIRAS~CIRB spectrum, our model slightly 
but significantly underestimates the source counts at 850 $\mu$m.
We should consider the discrepancy found in our model predictions
and detected sub-mm source counts, because in contrast, our model 
successfully reproduced the FIRAS~CIRB sub-mm spectrum.
Since, in our estimate, IR galaxies are fewer at high redshift compared 
with sub-mm-weighted model (e.g.\ Tan et~al.~1999), this result seems
qualitatively natural.

Here we consider the difficulty in measuring a source flux from noisy data.
Hogg, Turner (1998) clearly summarize and discuss the upward bias of the flux 
estimation in the study of source number counts.
Very recently Hogg (2000) extensively examined the confusion effect 
on the astrometric and photometric measurements.
Hogg showed an important result that if the observed number counts
has a steep slope like IR and sub-mm source counts, flux estimation
must suffer from serious confusion noise, and most of the faintest sources
can be spurious.
According to Hogg's discussion, the sampling strategies adopted in
the field of sub-mm survey are often insufficient.
Thus we do not make attempts to search the solution which satisfy the 
present sub-mm source counts in this study.
In order to settle down this problem, a much better rule-of-thumb for
the same flux limit level for the confusion is required.
Eales et~al.~(2000) also performed a Monte Carlo simulation of the source
extraction from the sub-mm images, and clearly showed that 
the estimated source fluxes are boosted upwards significantly, by the heavy 
confusion.

We discuss another possible source of the sub-mm discrepancy.
The hot gas in X-ray cluster of galaxies produces a spectral distortion
of the CMB: the Sunyaev--Zel'dovich (SZ) effect (Sunyaev, Zel'dovich 1972).
At wavelengths shorter than 1.38~mm, the SZ effect appears as a positive
source in the CMB and has a maximum intensity at $850\;\mu$m.
The number count and redshift distribution of clusters are 
actively studied by many authors (e.g. Barbosa et al.~1996; 
Kitayama, Sasaki, Suto 1998).
Kitayama et~al.~(1998) prediced the sub-mm cluster number counts based on 
the Press--Schechter formalism (Press, Schechter 1974).
The contribution of SZ clusters to the sub-mm numbers count is not 
negligible compared with sub-mm galaxy counts, while their contribution 
to the CIRB is small (Kitayama et al.~1998).
In addition, SZ cluster is a diffuse source, and thus they might affect
the flux estimation of sub-mm point sources in the cluster regions.
Hughes, Gatzta\~{n}aga (2000) has already included the effect of the 
SZ clusters in their sub-mm sky image simulations.

In addition, we should take the possible AGN contribution to
the sub-mm source counts into account, though there are some observational
constraints on their fraction in the sources (e.g. Haarsma, Partridge 1998).

Instruments with a much better angular resolution in sub-mm wavelengths, 
like forthcoming Large Millimeter and Submillimeter Array (LMSA) and further,
the international collboration of large arrays, Atacama Large Millimeter Array
(ALMA) will help to solve the confusion problem, 
and serve much information of the high-redshift status of galaxies.
We stress that large-area surveys targeted at bright submillimeter 
sources are also necessary to fix the evolution at $1 \ltsim z \ltsim 2$ 
(for the detailed examination of this issue, see Takeuchi et~al.~2000c and 
Hughes 2000).

\section{Discussion}

\subsection{Implication: High or Low Redshift?}

Our analysis shows that the steep slope of the IR number counts and
the peak intensity of the CIRB is attributed to relatively low-redshift 
($0.5 < z < 1$) galaxies.
Previous studies of the redshift distribution of IR galaxies based on 
the CIRB spectrum and IR number counts tend to claim that the IR -- sub-mm
observations support a model of high star-formation rate (SFR) at large 
redshift ($z > 2$), and consequently, imply a large number of heavily 
obscured distant galaxies.
We focus on this issue and examine our results in this subsection.

First we took care of the CIRB intensity at $140\;\mu$m (Hauser et~al.~1998).
As we discussed in Section~3, the observation and calibration of {\sl COBE}
DIRBE at this wavelength is reliable enough.
Therefore it seems strange that most of the previous studies completely 
ignored this observed point but focused on the other points.
Moreover, some previous model predictions failed to reproduce the overall 
spectral shape of the CIRB.
If we use normal dust temperature $T_{\rm dust} = 30 \sim 40$~K, the spectral 
shape of the CIRB naturally requires a significant contribution of
low-$z$ IR galaxies.

Gispert, Lagache, Puget (2000) coherently analysed the cosmic 
background from optical to millimeter wavelengths and derived the 
cosmic SFH, and also found a rapid increase of cosmic luminosity 
density, and consequently, the cosmic SFR density.
They also found that the contribution of the low-$z$ IR galaxies 
to the CIRB around the peak is almost 100\%.
Their evolution of the cosmic luminosity density at $z < 1$ is quite
consistent with our result.
Eales et al.\ (1999) studied their sub-mm survey and presented a similar 
SFH, although their peak of the SFH is located on slightly higher redshift.

Recently Scott et al.~(2000) performed a sub-mm follow-up observation of 
10 known FIRBACK 170-$\mu$m sources and reported some detections at 450 and 
$850\;\mu$m.
They found that the sub-mm fluxes of the FIRBACK sources are fairly weak.
We see, in their Figure~1, that the SEDs of these sources are consistent 
with those of IR galaxies at $0 < z < 0.5$.
Scott et~al.~(2000) mentioned that there remains a degree of freedom to 
hypothesize the very hot $T_{\rm dust}$.
They carefully considered another redshift estimation, variation of dust 
emissivity and dust temperature, and concluded that these sources have 
redshifts of $0 < z < 1.5$, which well agrees our result.

Juvela et al.~(2000) reported a result of galaxy count as a part of their 
ISOPHOT CIRB project.
They performed multiband surveys at wavelengths $90\;\mu$m, $150\;\mu$m, and
$180\;\mu$m, and relatively high source surface density suggested strong
galaxy evolution.
They compared the observed SEDs of the newly detected {\sl ISO}~sources with
the IR-luminous galaxy SEDs of Lisenfeld, Isaak, Hills (2000), 
and found that the SEDs of the detected ojbects are consistent with 
those of dusy galaxies at redshift $0.5 \ltsim z \ltsim 1$.
This also supports our estimated evolution of IR galaxies.

We must note that, in addition, the density parameter of the Universe 
$\Omega_0 = 1$ often assumed in many previous models.
Large $\Omega_0$ suppress the faintest slope of the galaxy number count and 
intensity of the CIRB (e.g.\ Takeuchi et al. 1999).
Therefore, a group of the models which prefer very high redshift IR 
galaxies can be reconciled with the limits of the FIRAS CIRB spectrum by
the large cosmic density.
On the other hand, the cosmological constant $\lambda_0$ does not 
significantly affect the model predictions of the CIRB and IR galaxy 
number counts, compared with the effect of $\Omega_0$.
When we use the low-$\Omega_0$ cosmology ($0.2 < \Omega < 0.3$), 
which is now regarded as highly plausible, the CIRB spectrum turns out to
be a meaningful observational constraint of galaxies at high redshift.

\subsection{Cosmic Star Formation History}

\begin{figure}[t]
\epsfxsize=7.0cm
\centerline{\epsfbox{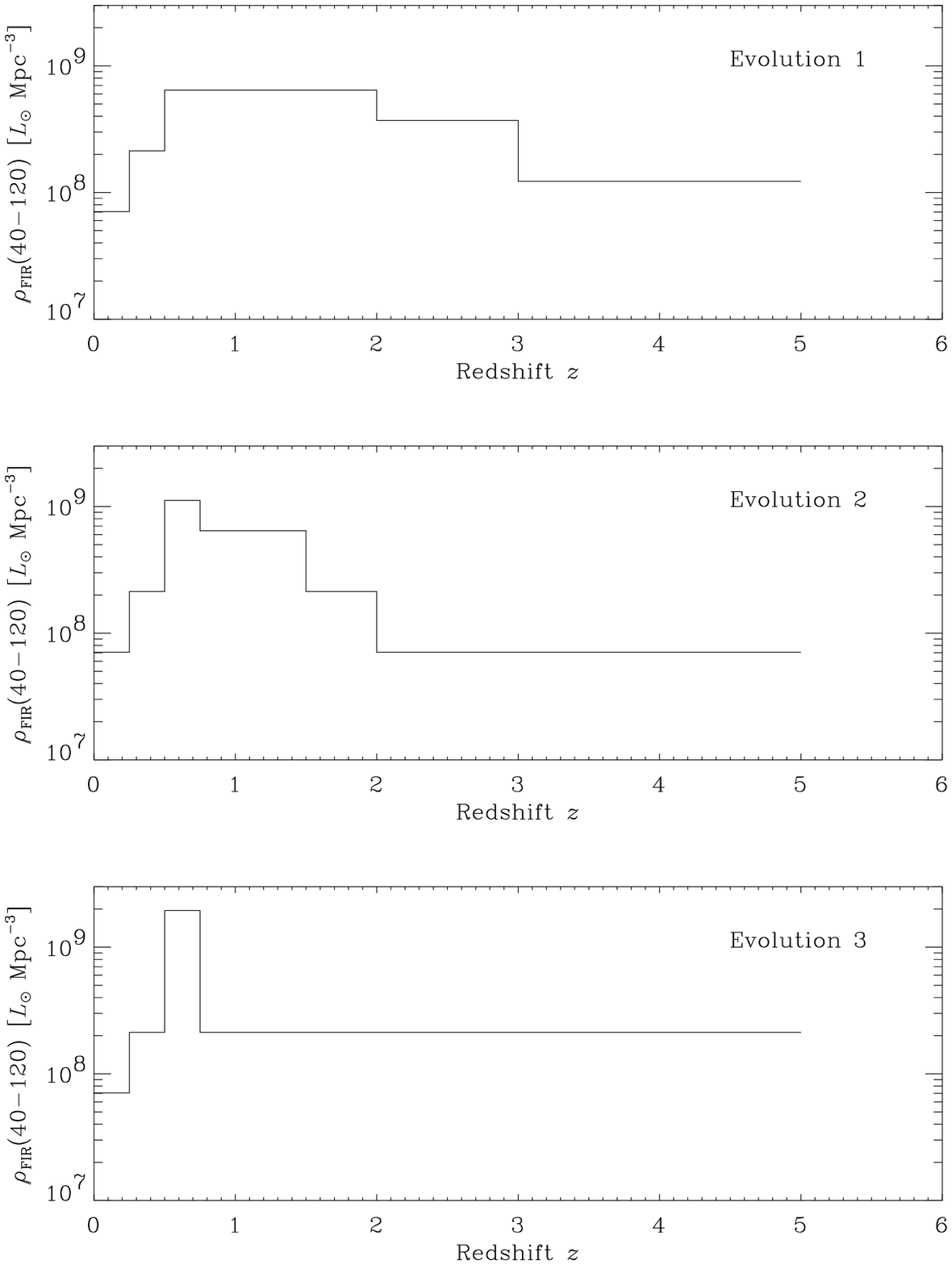}}
\caption
{Fig.\ 9.\ ---
  The evolution of the far-infrared luminosity density within
  the wavelength range from $40\;\mu$m to $120\;\mu$m, 
  $\rho_{\rm FIR}\mbox{(40--120)}$.
  The integrated luminosity density is derived from the monochromatic 
  luminosity density in two {\sl IRAS}~wavebands, $\rho_{60\,\mu{\rm m}}$ and
  $\rho_{100\,\mu{\rm m}}$ with formula of Helou et~al.~(1988).
}
\end{figure}

In this subsection we discuss the star formation rate per comoving 
density as a function of redshift. 
This is generally called the cosmic star formation history (SFH). 
Since, as we mentioned in Section~1,  the FIR luminosity of a galaxy is 
related to its star formation rate (SFR; e.g.\ Kennicutt 1998), we make 
attempts to convert the FIR luminosity density derived in the previous 
sections to the SFR density in the Universe. 

First, we estimate the comoving luminosity density in the FIR range of 
40--120 $\mu$m, $\rho_{\rm FIR}$(40--120). 
Using the well-accepted formula presented in the Appendix of 
Helou et~al.\ (1988), $\rho_{\rm FIR}$(40--120) is approximated as
\begin{eqnarray}
  &&\hspace{-13mm}\rho_{\rm FIR}(40\mbox{--}120) =
  3.26\times 10^{-19} \times \nonumber \\
  &&[2.58\rho_\nu (60)+1.00\rho_\nu (100)] \hspace{2mm}
  [L_\odot~{\rm Mpc^{-3}}],\label{eq:helou2}
\end{eqnarray}
where $\rho_\nu (60)$ and $\rho_\nu (100)$ are the comoving
luminosity density per unit frequency at 60 $\mu$m and 100 $\mu$m
$[{\rm erg\;s^{-1}Hz^{-1}Mpc^{-3}}]$, respectively. 
Helou et~al.\ (1988) asserted that the relation in their Appendix is 
valid in the range of dust temperature $T_{\rm d}$ and emissivity index 
$\gamma$ expected for a galaxy.
Thus, we expect that equation (\ref{eq:helou2}) is applicable
to the FIR luminosity density of the Universe, since the FIR emission is 
originating from dust in galaxies.
We show the history of $\rho_{\rm FIR}$(40--120) in Figure~9. 
Here we note that both $\rho_\nu (60)$ and $\rho_\nu (100)$ at redshift
$z$ are calculated from the LF and SEDs shown in Section~2.

Next, $\rho_{\rm FIR}$(40--120) is converted to the SFR per unit comoving 
volume, $\rho_{\rm SFR}$.
Recently, Inoue, Hirashita, Kamaya (2000) derived the conversion formula 
from FIR luminosity to the SFR.
Their formula is applied to the relation between $\rho_{\rm FIR}$(40--120) 
and $\rho_{\rm SFR}$ as
\begin{eqnarray}
  &&\hspace{-13mm}\rho_{\rm SFR}~[M_\odot~{\rm yr}^{-1}~{\rm Mpc}^{-3}] =
  \frac{2.4\times 10^{-10}(1-\eta )}{0.4-0.2f+0.6\epsilon}\, \nonumber \\
  &&\hspace{1.5cm}
  \times\rho_{\rm FIR}(40\mbox{--}120)\quad [L_\odot~{\rm Mpc}^{-3}]\; ,
\end{eqnarray}
where $f$ is the fraction of ionizing photons absorbed by neutral hydrogen 
(i.e.\ $(1-f)$ is the fraction of ionizing photons absorbed by dust grains), 
$\epsilon$ is the efficiency of dust absorption for nonionizing photons, 
and $\eta$ is the cirrus fraction of observed FIR luminosity.
Since Inoue et al.'s expression is based on the luminosity in the range of 
8--1000 $\mu$m, we have converted it to the luminosity in 40--120 $\mu$m 
by the factor of 1.4 (e.g.\ Buat, Xu 1995).
For convenience of discussions, we define the factor $C_{\rm FIR}$ as
\begin{eqnarray}
  C_{\rm FIR}\; [M_\odot~L_\odot^{-1}~{\rm yr}^{-1}] =
  \frac{2.4\times 10^{-10}(1-\eta )}{0.4-0.2f+0.6\epsilon}\;,
\end{eqnarray}
or equivalently
\begin{eqnarray}
  &&\hspace{-10mm}\rho_{\rm SFR}\; [M_\odot~{\rm yr}^{-1}~{\rm Mpc}^{-3}] = 
  \nonumber \\
  &&\hspace{1.3cm}C_{\rm FIR} \cdot \rho_{\rm FIR}(40\mbox{--}120)\; 
  [L_\odot~{\rm Mpc}^{-3}]\;.
\end{eqnarray}

\begin{figure}[t]
\epsfxsize=7.0cm
\centerline{\epsfbox{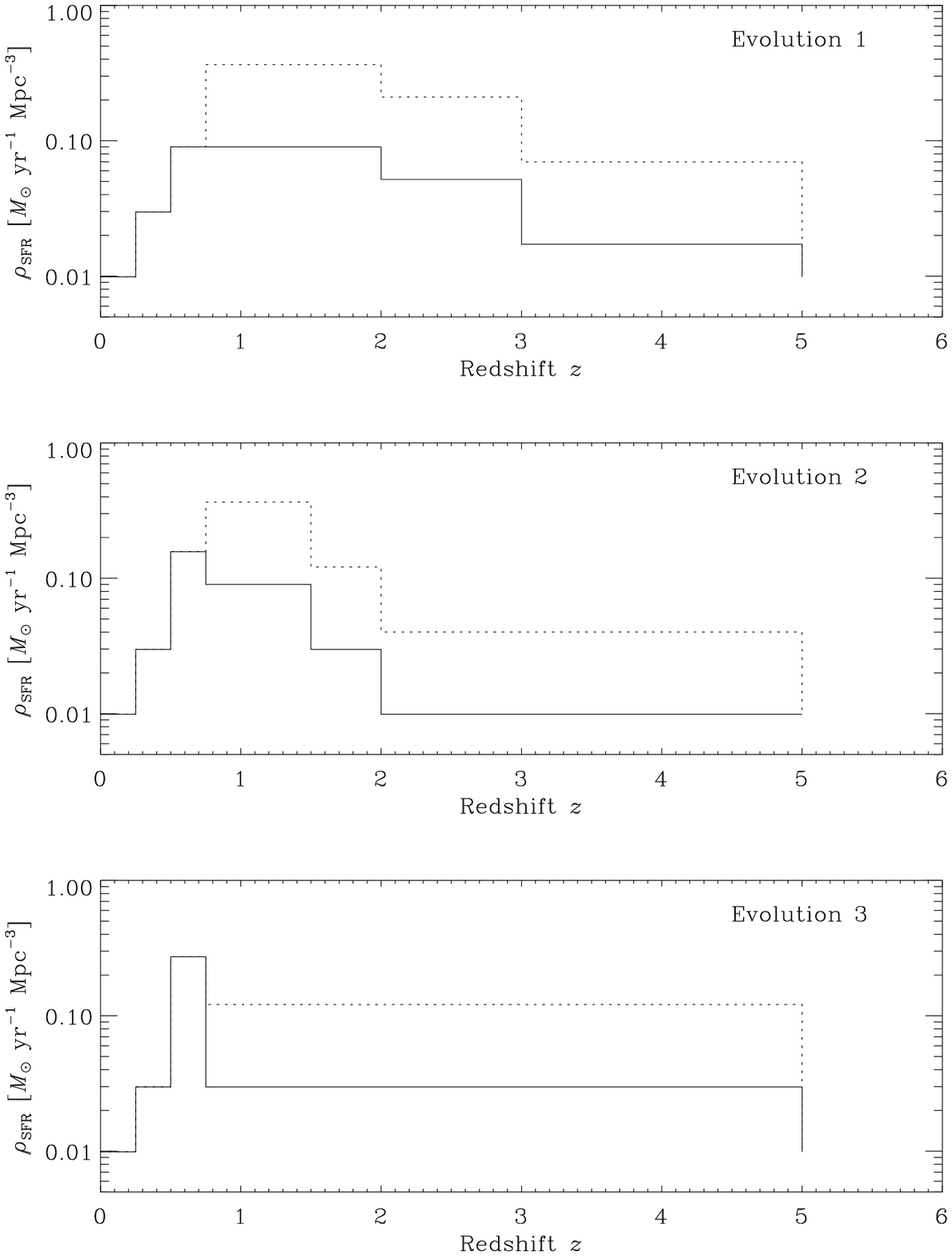}}
\caption
{Fig.\ 10.\ ---
  The star formation history derived form cosmic infrared background 
  and IR number counts. 
  The solid line represents the case where the factor $C_{\rm FIR}$ 
  for the conversion from FIR luminosity to SFR is fixed at the value
  corresponding to the solar metallicity. The dotted line presents the
  case $C_{\rm FIR}$ is 4.1 times larger because of the
  inefficient conversion from stellar luminosity to FIR luminosity.
}
\end{figure}

The above parameters $f$ and $\epsilon$ depend on dust-to-gas ratio, 
which is related to metallicity (e.g. Hirashita 1999). 
Thus, $C_{\rm FIR}$ depends on metallicity (Hirashita, Inoue, Kamaya 2000) 
and it varies as the metallicity evolution of galaxies. 
As the metallicity evolves from 0.01 to 1 times the solar value, 
$C_{\rm FIR}$ changes from 
$5.8\times 10^{-10}~M_\odot~L_\odot^{-1}~{\rm yr}^{-1}$
to $1.4\times 10^{-10}~M_\odot~L_\odot^{-1}~{\rm yr}^{-1}$.
This is because the conversion efficiency from the stellar luminosity to 
the FIR luminosity increases as the dust-to-gas ratio (metallicity) becomes 
large.
For simplicity, the cirrus fraction is set constant as $\eta =0.5$, since 
as stated in Hirashita et al.~(2000), it is difficult to relate $\eta$ to 
the metallicity from our present knowledge.

In Figure~10, we show the derived cosmic SFH.
The solid line indicates the history determined by using constant 
$C_{\rm FIR}$, where the value for the solar metallicity is adopted (i.e., 
$C_{\rm FIR} = 1.4\times 10^{-10}\;[M_\odot~L_\odot^{-1}~{\rm yr}^{-1}]$).
The dotted line represents the history by using
$C_{\rm FIR} = 1.4\times 10^{-10}\;[M_\odot~L_\odot^{-1}~{\rm yr}^{-1}]$
for $z<1$, while
$C_{\rm FIR}=5.8\times 10^{-10}\;[M_\odot~L_\odot^{-1}~{\rm yr}^{-1}]$
for $z>1$ to examine the extremely metal-poor case. 
In order to be consistent, we should model the cosmic chemical evolution. 
However, since the model produces further undetermined parameters, 
we do not treat such a model in this paper. 
We left detailed and consistent modeling to the future works (for reference 
on the cosmic chemical evolution, see e.g., Pei, Fall, Hauser 1999). 
We should also decrease $\eta$ as the FIR luminosity density increases, 
since the contribution from the cirrus component becomes smaller as 
the infrared luminosity becomes larger (e.g., Beichman, Helou 1991).
If $\eta$ is set as 0 for an extreme case, $C_{\rm FIR}$ becomes two
times larger than that estimated above.
Accordingly, the uncertainty of the $C_{\rm FIR}$ induced by the cirrus 
fraction is estimated to be a factor of two.

In a realistic situation, we expect that the cosmic SFH lies between these 
two lines in Figure~10.
In summary, we conclude that $\rho_{\rm SFR}$ increases to $\sim 10^{-1} \; 
M_\odot \, {\rm yr}^{-1} {\rm Mpc}^{-3}$ at $z \sim 1$ and remains almost 
constant or declines gradually for $z > 1$, very similar to the SFH presented 
in Figure~9 of Steidel et al.\ (1999) and that in Figure~6 of 
Gispert et al.~(2000).

\section{Summary and Conclusion}

We estimated the galaxy evolution from infrared (IR) galaxy number count
and cosmic infrared background (CIRB).
We constructed an empirical galaxy count model based on the spectral energy 
distributions (SEDs) and luminosity function (LF), both of which are
constructed from the {\sl IRAS}~results.
The SEDs are made based on the {\sl IRAS}~color--luminosity relations and 
a tight far-IR--radio correlation.
Using this model we estimated the plausible evolution of galaxies from 
the data in a nonparametric way.

We found that a violent evolution of galaxies as a whole in  relatively
low redshifts ($z \ltsim 0.5 \mbox{--} 0.75$) should exist to explain the 
observed number count of galaxies in the IR wavelengths and the CIRB.
Thus we conclude that the forthcoming galaxy surveys in the IR by
future facilities (e.g. ASTRO-F, SOFIA, {\sl SIRTF}, {\sl FIRST}, and 
{\sl NGST}) have a crucial importance to understand the global history of 
galaxies, since the IR wavelength is suitable for the investigations of 
the redshift range $0 < z < 1$.

On the other hand, at present, star formation history at high redshift
($z > 1$) requires further observations and examinations of submillimeter 
(sub-mm) sources.
In addition, source confusion severely affect the flux estimation in
sub-mm.
Instruments with much better angular resolution in sub-mm wavelengths, 
like forthcoming LMSA and ALMA will help to solve the confusion problem, 
and serve much information on the high-redshift status of galaxies.
We stress that large-area surveys targeted at bright sub-mm
sources are also necessary to fix the evolution at $z \gtsim 1 - 2$.

Then we converted the FIR luminosity density to the SFR density in the 
Universe. 
We estimated the comoving luminosity density in the FIR range of 
40--120 $\mu$m by the approximate formula of Helou et~al.\ (1988),
and $\rho_{\rm FIR}$(40--120) is converted to the star formation rate 
per unit comoving volume, $\rho_{\rm SFR}$.
We used the conversion formula from FIR luminosity to the SFR derived by
Inoue et al.~(2000), which took the following three 
parameters into account: 
the fraction of ionizing photons absorbed by neutral hydrogen $f$, 
the efficiency of dust absorption for nonionizing photons $\epsilon$, 
and the cirrus fraction of the observed far-infrared luminosity $\eta$.
Since $f$ and $\epsilon$ depend on dust-to-gas ratio which is related to 
the metallicity, we can include the effect of the metal evolution in galaxies.
We observe that the SFR density increases to $\sim 10^{-1} \; M_\odot \, 
{\rm yr}^{-1} {\rm Mpc}^{-3}$ at $z \sim 1$ and it remains almost constant 
for $z > 1$, very similar to the star formation history presented by 
Steidel et al.\ (1999) and Gispert et al.\ (2000).

\vspace{1pc}\par
First we thank the anonymous referee for helpful comments, which improved
the quality of the paper.
We wish to acknowledge Dr.~Hiroshi Shibai, Dr.~Mitsunobu Kawada, 
Dr.~Hidenori Takahashi, Dr.~Chris P.\ Pearson, Dr.~Hiroshi Matsuo, Dr.~Tetsu
Kitayama, and Dr.\ T.~N.~Rengarajan for helpful discussions and suggestions.
HH and KY acknowledge the Research Fellowships of the Japan Society for 
the Promotion of Science for Young Scientists.
We made extensive use of the NASA's Astrophysics Data System Abstract 
Service (ADS).

\section*{Appendix.\ Performance of the {\sl IRIS}~Far Infrared All-sky Survey}

\setcounter{equation}{0}
\renewcommand{\theequation}{A\arabic{equation}}

In this Appendix we present the performance of the Japanese ASTRO-F 
({\sl IRIS}) all-sky survey.
The bandpass system of the FIR instrument, ASTRO-F Far Infrared Surveyor 
(FIS) consists of two narrow bands, N60 ($50 - 70\;\mu$m) and N170 
($150 - 200\;\mu$m), and two wide bands, WIDE-S ($50 - 110\;\mu$m) and 
WIDE-L ($110 - 200\;\mu$m).
We show the expected number counts at each band in Figures~11 and 12.
The vertical dot-dot-dot-dashed line presents the $5\sigma$-flux detection 
limit in each band.
The detection limits are estimated as 39~mJy and 110~mJy for N60 and N170, 
and 16~mJy and 90~mJy for WIDE-S and WIDE-L, respectively (Takahashi 
et~al.\ 2000). 
In order to obtain realistic galaxy counts, the wide bandpass wavelength 
range is involved in calculation.
Figures~11 and 12 show that $\mbox{several} \times 10^5$ galaxies per sr 
will be detected by the sets of narrow bandpasses, and roughly $10^6$ galaxies
per sr will be detected in WIDE-S and -L.

\begin{figure*}[t]
\epsfxsize=12.0cm
\centerline{\epsfbox{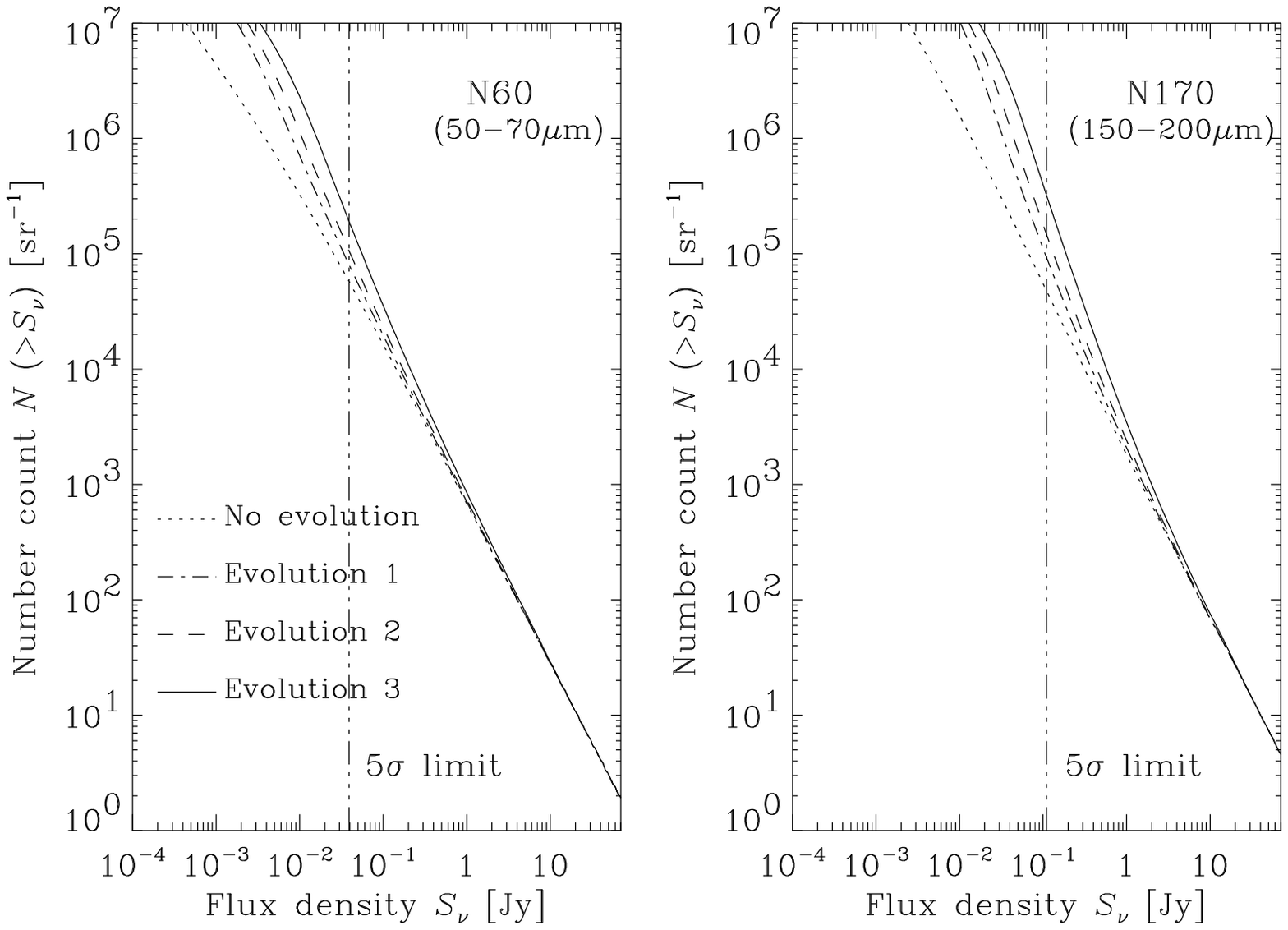}}
\vspace{-2cm}
\caption
{Fig.\ 11.\ ---
The expected number counts in the ASTRO-F far-infrared 
    all-sky galaxy survey at two narrow band filters, called N60 and N170.
    The wavelength ranges are $50 - 70 \; \mu$m (N60) and 
    $150 - 200 \; \mu$m (N170).
}
\end{figure*}
\begin{figure*}[t]
\epsfxsize=12.0cm
\centerline{\epsfbox{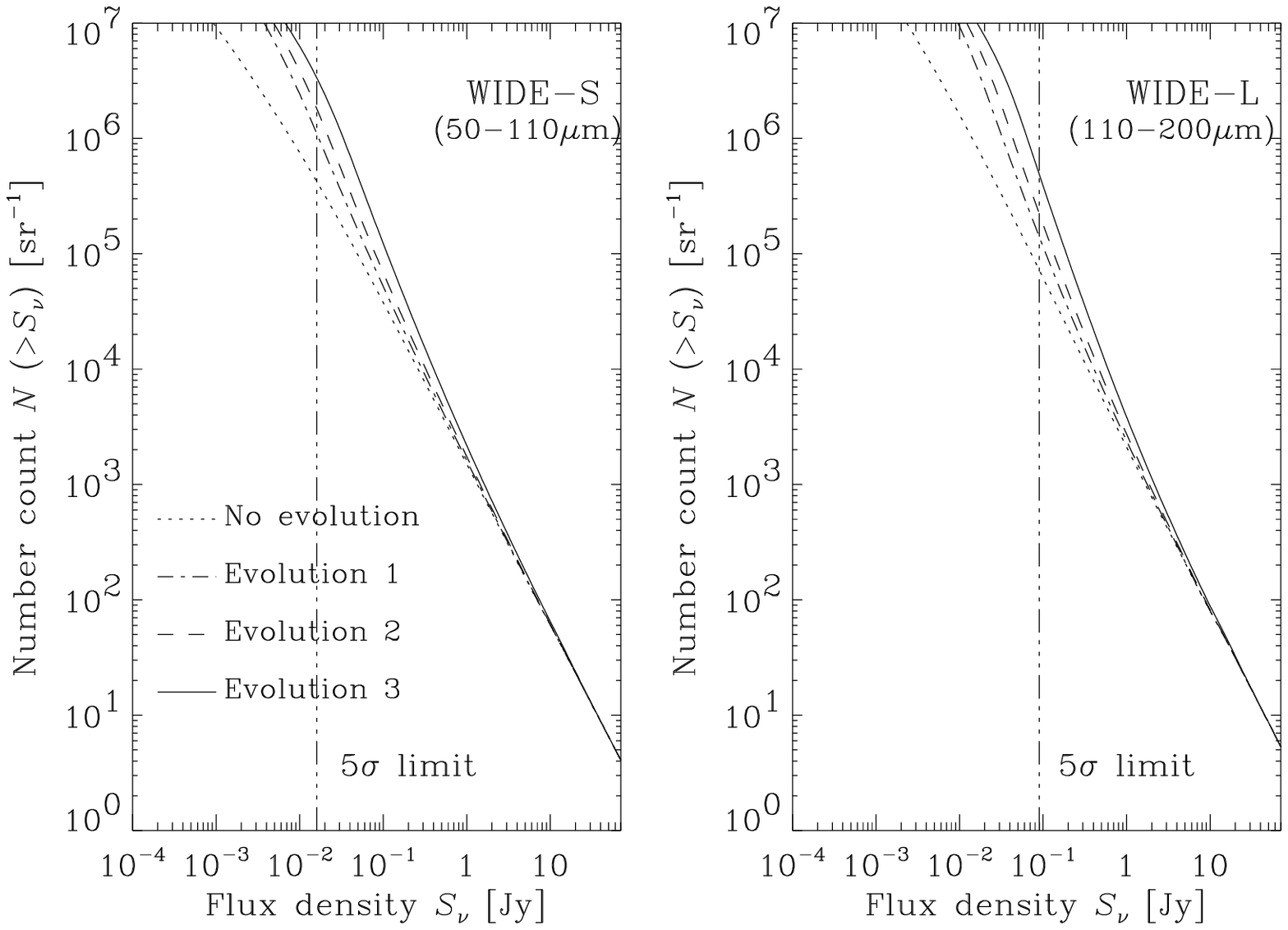}}
\vspace{-2cm}
\caption
{Fig.\ 12.\ ---
The expected number counts in the ASTRO-F far-infrared 
  all-sky galaxy survey at two wide band filters, called WIDE-S and WIDE-L.
  The wavelength ranges are $50 - 110 \; \mu$m (WIDE-S) and 
  $110 - 200 \; \mu$m (WIDE-L).
}
\end{figure*} 

In order to determine the effect of galaxy evolution from survey data, 
a significant sky area should be scanned, to suppress the variation of
galaxy surface density.
Since galaxies are clustered on the sky, the nominal error bar estimated from 
the Poisson assumption is an underestimation, and we must take the galaxy 
angular correlation function into account.
Consider a survey area $\Omega$, and divide it into small cells 
$\{ \Delta\Omega_i \}$ so that the number of galaxies in the cell
$\{ \Delta\Omega_i \}$, ${\cal N}_i = 0$ or 1.
We set the mean galaxy surface number density ${\cal N}$.
Then we have a mean number in a solid-angle cell $\Delta\Omega_i$, 
${\cal N}_i$, 
\begin{eqnarray}
  \langle {\cal N}_i \rangle = {\cal N} \Delta\Omega_i\;.
\end{eqnarray}
By definition, we have $\langle {\cal N}_i \rangle = 
\langle {\cal N}_i^2 \rangle =   \langle {\cal N}_i^3 \rangle = \cdots$.
We observe
\begin{eqnarray}
  &&\hspace{-1.2cm}\langle {\cal N}_i {\cal N}_j \rangle =
  {\cal N}^2 (1 + w(\theta_{ij}))\,\Delta\Omega_i\Delta\Omega_j \\
  &&\hspace{-1.2cm}\langle ({\cal N}_i - \langle {\cal N}_i \rangle)
  ({\cal N}_j - \langle {\cal N}_j \rangle) \rangle \nonumber \\
  &&= \langle {\cal N}_i {\cal N}_j \rangle - 
    \langle {\cal N}_i \rangle \langle {\cal N}_j \rangle \nonumber \\
    &&= {\cal N}^2 (1 + w(\theta_{ij}))\,\Delta\Omega_i\Delta\Omega_j 
    - {\cal N}^2 \,\Delta\Omega_i\Delta\Omega_j \nonumber \\
    &&= {\cal N}^2 w(\theta_{ij})\,\Delta\Omega_i\Delta\Omega_j
\end{eqnarray}
where $w(\theta)$ is the angular two-point correlation function of
galaxies.
Next we consider the number of galaxies in the survey area $\Omega$, $N$.
We have
\begin{eqnarray}
  &&\hspace{-1.2cm}\langle N \rangle =
  \sum_{\Delta\Omega_i \subset \Omega} \langle {\cal N}_i \rangle = 
  \int_\Omega {\cal N} {\rm d} \Omega = {\cal N}\Omega\;,  \\
  &&\hspace{-1.2cm}\langle N^2 \rangle =
  \langle \sum_{i} {\cal N}_i \sum_{j} {\cal N}_j \rangle = 
  \sum_{i} \langle {\cal N}^2_i \rangle + 
  \sum_{i \neq j} \langle {\cal N}_i {\cal N}_j \rangle \nonumber \\
  &&\hspace{-0.33cm} = \sum_{i} \langle {\cal N}_i \rangle + 
  \sum_{i \neq j} \langle {\cal N}_i {\cal N}_j \rangle \nonumber \\
  &&\hspace{-0.33cm} = \int_\Omega {\cal N} {\rm d} \Omega + 
  \int \int_\Omega {\cal N}^2 (1 + w(\theta_{12}))\,{\rm d} 
  \Omega_1 {\rm d} \Omega_2 \nonumber \\
  &&\hspace{-0.33cm} = {\cal N}\Omega + {\cal N}^2\Omega^2 + 
  {\cal N}^2 \int \int_\Omega w(\theta_{12})\,{\rm d} 
  \Omega_1 {\rm d} \Omega_2\;.
\end{eqnarray}
and thus 
\begin{eqnarray}
  \hspace{-3mm}\langle (N - \langle N \rangle)^2 \rangle 
  \hspace{-0.3cm}&=&\hspace{-0.3cm}
  \langle N^2 \rangle - \langle N \rangle^2 \nonumber \\
  \hspace{-3mm}\hspace{-0.3cm}&=&\hspace{-0.3cm} {\cal N}\Omega + 
  {\cal N}^2 \int \int_\Omega w(\theta_{12})\,{\rm d} 
  \Omega_1 {\rm d} \Omega_2\;.
\end{eqnarray}
If we assume that we take a sufficiently large area than the coherent
scale of angular clustering, we can approximate the above expression as
\begin{eqnarray}
  \langle (N - \langle N \rangle)^2 \rangle 
  \hspace{-0.3cm}&\simeq&\hspace{-0.3cm}
  {\cal N}\Omega + {\cal N}^2 \Omega \int_\Omega w(\theta)\,{\rm d} \Omega
  \nonumber \\
  \hspace{-0.3cm}&=&\hspace{-0.3cm} {\cal N} \Omega 
  \left( 1 + {\cal N} \int_\Omega w(\theta)\,{\rm d} \Omega \right)\;.
\end{eqnarray}
Here we evaluate the `signal-to-noise ratio' of the number count 
${\sf S/N}$:
\begin{eqnarray}
  {\sf S/N} \hspace{-0.3cm}&\equiv&\hspace{-0.3cm} \frac{\langle N \rangle}{
    \sqrt{\langle (N - \langle N \rangle)^2 \rangle}} \nonumber \\
  \hspace{-0.3cm}&\simeq&\hspace{-0.3cm}
  \frac{{\cal N}\Omega}{\sqrt{{\cal N}\Omega
      \left( 1 + {\cal N} \int_\Omega w(\theta)\,{\rm d} \Omega \right)}} 
  \nonumber \\
  \hspace{-0.3cm}&\simeq&\hspace{-0.3cm} \sqrt{\frac{\Omega}{
      \int_\Omega w(\theta)\,{\rm d} \Omega }}\;.
\end{eqnarray}
We see that the ${\sf S/N}$ depends on the angular correlation strength and
the solid angle of the survey, and is almost independent of the surface 
density of the sources ${\cal N}$.
This fact shows that, in order to determine galaxy evolution from 
number counts, large-area survey is substantially required.
The angular correlation function of the {\sl IRAS}~galaxies is $w(\theta) =
(\theta/\theta_0)^{-0.66}$ ($\theta_0 = 0.\hspace{-3pt}{}^\circ11$: 
Lahav, Nemiroff, Piran 1990).
The angular correlation of galaxies detected by the ASTRO-F survey is
obtained by using the scaling relation of $w(\theta)$ with detection limit,
through relativistic Limber's equation (e.g. Peebles 1980).
If we would like ${\sf S/N} = 10$, a survey with $\sim 50 \;\mbox{--} \;300 
\; \mbox{deg}^2$ is required, depending on the considered flux.
We see that, although the deeper is desirable, but 
{\sl the wider is necessary}.
The ASTRO-F all-sky survey promises a unique opportunity to execute such
a large-area survey.
Thus we can expect for ASTRO-F survey to distinguish between Evolutions~1, 
2, and 3 clearly, and to fix the evolutionary history of IR galaxies up to 
$z \sim 1$.

\section*{References} \vspace{1mm}
\re
Allamandola L.\ J., Tielens, G.\ G.\ M., Barker, J.\ R.\ 1989, ApJS, 71, 733
\re 
Altieri, B., et al.\ 1999, A\&A, 343, L65
\re
Ashby, M.\ L.\ N., Hacking, P.\ B., Houck, J.\ R., Soifer, B.\ T., 
Weisstein, E.\ W.\ 1996, ApJ, 456, 428
\re 
Aussel, H., C\'{e}sarsky, C., Elbaz, D., Starck, J.\ L.\ 1999, A\&A, 342, 313
\re
Barbosa, D., Bartlett, J.\ G., Blanchard, A., Oukbir, J.\ 1996, A\&A, 314, 13
\re 
Barger, A.\ J., Cowie, L.\ L., Sanders, D.\ B., Fulton, E., Taniguchi, Y., 
Sato, Y., Kawara, K., Okuda, H.\ 1998, Nature, 394, 248
\re 
Barger, A.\ J., Cowie, L.\ L., Sanders, D.\ B.\ 1999a, ApJ, 518, L5
\re
Barger, A.\ J., Cowie, L.\ L., Smail, I., Ivison, R.\ J., Blain, A.\ W., 
Kneib, J.\ -P.\ 1999b, AJ, 117, 2656
\re
Beichman, C.\ A., Helou, G.\ 1991, ApJ, 370, L1
\re
Bertin, E., Dennefeld, M., Moshir, M.\ 1997, A\&A, 323, 685
\re 
Blain, A.\ W., Kneib, J.\ -P., Ivison, R.\ J., Smail, I.\ 1999, ApJ, 512, L87
\re
Blain, A.\ W., Ivison, R.\ J., Kneib, J.\ -P., Smail, I.\ 2000, in The 
Hy-Redshift Universe, ed~A.\ J.\ Bunker, W.\ J.\ M.\ van Breugel, 
ASP Conference vol.193, ASP: San Francisco, p.246
\re
Bregman, J.\ N., Hogg, D.\ E., Roberts, M.\ S.\ 1992, ApJ, 387, 484
\re
Buat, V.\ Xu, C.\ 1995, A\&A, 293, L65
\re 
Clements, D.\ L., D\'{e}sert, F.-X., Franceschini, A., Reach, W.\ T., 
Baker, A.\ C., Davies, J.\ K., C\'{e}sarsky, C.\ 1999, A\&A, 346, 383
\re
Condon J.\ J.\ 1992, ARA\&A, 30, 575
\re 
Dole, H.\ et al.\ 2000, in {\sl ISO}~Surveys of a Dusty Universe, 
ed D.\ Lemke, M.\ Stickel, K.\ Wilke, p.54
\re 
Dwek, E.\ et al.\ 1997, ApJ, 475, 565
\re
Dwek, E.\ et al.\ 1998, ApJ, 508, 106
\re 
Eales, S., Lilly, S., Gear, W., Dunne, L., Bond, J.\ R., Hammer, F., 
Le F\`{e}vre, O., Crampton, D.\ 1999, ApJ, 515, 518
\re
Eales, S., Lilly, S., Webb, T., Dunne, L., Gear, W., Clements, D., Yun, M.
2000, ApJ, in press, astro-ph/0009154
\re
Elbaz, D.\ et al.\ 1999, A\&A, 351, L37
\re 
Fixsen, D.\ J., Dwek, E., Mather, J.\ C., Bennet, C.\ L.\ Shafer, R.\ A.
1998, ApJ, 508, 123,
\re
Flores, H., et al.\ 1999a, A\&A, 343, 389
\re 
Flores, H., Hammer, F., Thuan, T.\ X., C\'{e}sarsky, C., D\'{e}sert, F.-X., 
Omont, A., Lilly, S.\ J., Eales, S., Crampton, D., Le F\`{e}vre, O.\ 
1999b, ApJ, 517, 148
\re 
Gardner, J.P.\ 1998, PASP, 110, 291
\re
Gispert, R., Lagache, G., Puget, J.\ -L.\ 2000, A\&A, 360, 1
\re
Guiderdoni, B., Hivon, E., Bouchet, F.\ R., Maffei, B.\ 1998, MNRAS, 295, 877
\re
Haarsma, D.\ B., Partridge, R.\ B.\ 1998, ApJ, 503, L5
\re 
Hauser, M.\ G., et al.\ 1998, ApJ, 508, 25
\re
Helou, G., Soifer, B.\ T., Rowan-Robinson, M.\ 1985, ApJ, 298, L7
\re
Helou, G., Khan, I.\ R., Malek, L., Boehmer, L.\ 1988, ApJS, 68, 151
\re
Hirashita, H.\ 1999, ApJ, 510, L99
\re
Hirashita H., Takeuchi, T.\ T., Ohta, K., Shibai, H.\ 1999, PASJ, 51, 81
\re
Hirashita, H., Inoue, A.\ K., Kamaya, H.\ 2000, A\&A, submitted
\re
Hogg, D.\ W., Turner, E.\ L.\ 1998, PASP, 110, 727
\re
Hogg, D.\ W.\ 2000, AJ, submitted, astro-ph/0004054
\re 
Holland, W.\ S., et al.\ 1999, MNRAS, 303, 659
\re
Hughes, D.\ H.\ et al.\ 1998, Nature, 394, 241
\re
Hughes, D.\ H., Gazta\~{n}aga, E.\ 2000, in Star Formation from the Small to 
the Large Scale, ed F.\ Favata, A.\ A.\ Kaas, A.\ Wilson, ESA SP--445, p.29
\re
Hughes, D.\ H.\ 2000, in Clustering at High Redshift, ed\ A.~Mazure O.\ 
Le~F\`{e}vre, ASP Conf.\ Series, in press, astro-ph/0003414
\re
Inoue, A.\ K., Hirashita, H., Kamaya, H.\ 2000, PASJ, 52,539
\re
Ishii, T.\ T., Takeuchi, T.\ T., Hirashita, H., Yoshikawa, K.\ 2000, in
Star Formation from the Small to the Large Scale, ed F.\ Favata, A.\ A.\ 
Kaas, A.\ Wilson, ESA SP--445, p.421
\re
Juvela, M., Mattila, K., Lemke, D.\ 2000, A\&A, 360, 813
\re 
Kawara, K.\ et al.\ 1998, A\&A, 336, L9
\re 
Kawara, K.\ et al.\ 2000, 
in {\sl ISO}~Surveys of a Dusty Universe, ed D.\ Lemke, 
M.\ Stickel, K.\ Wilke, p.50
\re
Kennicutt, R.\ C.\ Jr.\ 1998, ARA\&A, 36, 189
\re
Kitayama, T., Sasaki, S., Suto, Y.\ 1998, PASJ, 50, 1
\re
Lagache, G., Abergel, A., Boulanger, F., D\'{e}sert, F.\ X., Puget, J.\ -L.
1999, A\&A, 344, 322
\re
Lagache, G., Haffner, L.\ M., Reynolds, R.\ J., Tufte, S.\ L.\ 2000, A\&A, 
354, 247
\re
Lahav, O., Nemiroff, R.\ J., Piran, T.\ 1990, ApJ, 350, 119
\re
Lisenfeld, U., Isaak, K.\ G., Hills, R.\ 2000, MNRAS, 312, 433
\re
Malkan, M.\ A., Stecker, F.\ W.\ 1998, ApJ, 496, 13
\re
Matsuhara, H., et al.\ 2000, A\&A, in press, astro-ph/0006444
\re 
Okuda, H.\ 2000, in {\sl ISO}~Surveys of a Dusty Universe, 
ed D.\ Lemke, M.\ Stickel, K.\ Wilke, p.40
\re
Okuda, H.\ et al.\ 2000, in preparation
\re
Oliver, S., et al.\ 1997, MNRAS, 289, 471
\re
Oliver, S., et al.\ 2000a, {\sl ISO}~Surveys of a Dusty Universe, ed 
D.\ Lemke, M.\ Stickel, K.\ Wilke, p.28
\re
Oliver, S., et al.\ 2000b, MNRAS, 316, 749
\re
Pearson, C.\ P., Rowan-Robinson, M.\ 1996, MNRAS, 283, 174
\re
Pearson, C.\ P., Matsuhara, H., Watarai, H., Matsumoto, T., Onaka, T.\ 
2000, MNRAS, submitted, astro-ph/0008472 
\re
Peebles, P.\ J.\ E.\ 1980, The Large-Scale Structure of the Universe, Princeton
University Press, Princeton
\re 
Pei, Y.\ C., Fall, S.\ M., Hauser, M.\ G.\ 1999, ApJ, 522, 604
\re
Puget, J.\ -L., Abergel, A., Bernard, J.\ -P., Boulanger, F., Burton, W.\ B.,
D\'{e}sert, F.\ -X., Hartmann, D.\ 1996, A\&A, 308, L5
\re
Puget, J.\ -L., et al.\ 1999, A\&A, 345, 29
\re 
Rowan-Robinson, M., Saunders, W., Lawrence, A., Leech, K.\ 
1991, MNRAS, 253, 485
\re 
Rush, B., Malkan, M.\ A., Spinoglio, L.\ 1993, ApJS, 89, 1
\re
Press, W.\ H.\ Schechter, P.\ L.\ 1974, ApJ, 187, 425
\re
Saunders, W., Rowan-Robinson, M., Lawrence, A., Efstathiou, G., 
Kaiser, N., Ellis, R. S., Frenk, C. S.\ 1990, MNRAS, 242, 318
\re
Scott, D., et al.\ 2000, A\&A, 357, L5
\re
Serjeant, S.\ et al.\ 2000, MNRAS, 316, 718
\re 
Shibai, H., Okumura, K., Onaka, T.\ 2000, in Star Formation 1999, 
ed.~T.\ Nakamoto, Nobeyama Radio Observatory:NRO, p.67
\re
Smail, I., Ivison, R.\ J., Blain, A.\ W.\ 1997, ApJ, 490, L5
\re 
Smith B.\ J., Kleinman, S.\ G., Huchra, J.\ P., Low, F.\ J.\ 1987, 
ApJ, 318, 161
\re 
Soifer B.\ T., Sanders, D.\ B., Madore, B.\ F., Neugebauer, G., 
Danielson, G.\ E., Elias, J.\ H., Lonsdale, C.\ J., Rice, W.\ L.\ 
1987, ApJ, 320, 238
\re 
Soifer B.\ T., Neugebauer, G.\ 1991, AJ, 101, 354
\re
Springel, V., White, S.\ D.\ M.\ 1998, MNRAS, 298, 143
\re
Steidel, C.\ C., Adelberger, K.\ L., Giavalisco, M., Dickinson, M.\ 
1999, ApJ, 519, 1
\re 
Stickel, M., et al.\ 1998, A\&A, 336, 116
\re
Sunyaev, R.\ A., Zel'dovich, Ya.\ B.\ 1972, Comm.\ Astrophys.\ Space Phys.\
4, 173
\re
Takahashi, H., et al.\ 2000, in UV, Optical, and IR Space Telescopes and 
Instruments, ed\ J.\ B.\ Breckinridge J.\ Jacobsen, Proc.\ SPIE, 4013, 47
\re 
Takeuchi T.\ T., Hirashita, H., Ohta, K., Hattori, T.\ G., Ishii, T.\ T., 
Shibai, H.\ 1999, PASP, 111, 288
\re
Takeuchi, T.\ T., Shibai, H., Ishii, T.\ T.\ 2000a, Adv.\ Space Res., 
submitted
\re
Takeuchi, T.\ T., Ishii, T.\ T., Hirashita, H., Yoshikawa, K., Mazmine, K.\ 
2000b, in Star Formation 1999, ed.\ T.\ Nakamoto, Nobeyama Radio 
Observatory:NRO, p.58 
\re
Takeuchi, T.\ T., Kawabe, R., Kohno, K., Nakanishi, K., Ishii, T.\ T., 
Hirashita, H., Yoshikawa, K.\ 2000c, PASP, submitted
\re
Tan, J.\ C., Silk, J., Balland, C.\ 1999, ApJ, 522, 579
\re
Xu, C.\ Hacking, P.\ B., Fang, F., Shupe, D.\ L., Lonsdale, C.\ J., 
Lu, N.\ Y., Helou, G., Stacey, G.\ J., Ashby, M.\ L.\ N.\ 1998, ApJ, 508, 576

\label{last}

\end{document}